\documentclass[aps,pra,onecolumn,nofootinbib,citeautoscript,10pt]{revtex4-2}  
\usepackage{amsmath,amssymb,bm, makecell} 

\usepackage{xcolor}
\usepackage{svg}

\usepackage{graphicx}

\usepackage[tight]{subfigure} 

\usepackage{amsmath}

\usepackage{color} 
\usepackage[papersize={8.5in,11in}]{geometry}
\usepackage[colorlinks=true]{hyperref}
\hypersetup{        
    unicode=false,          
    pdftoolbar=true,        
    pdfmenubar=true,        
    pdffitwindow=false,     
    pdfstartview={FitH},    
    pdfkeywords={keyword1} {key2} {key3}, 
    pdfnewwindow=true,      
    colorlinks=true,       
    linkcolor=magenta, 
    citecolor=blue,        
    filecolor=magenta,      
    urlcolor=blue           
} 

\usepackage{graphicx}
\usepackage{dcolumn}
\usepackage{color}
\usepackage{amssymb,amsmath}
\usepackage{tabularx,graphicx}
\usepackage{epstopdf}
\usepackage{latexsym}
\usepackage{colortbl}
\usepackage{psfrag}
\usepackage{bbm,bm,array,physics}
\usepackage{dsfont}

 \def\*#1{\mathbf{#1}} 

\newcommand{\be}{\mathbf{e}}

\def\be{\begin{eqnarray}}
\def\ee{\end{eqnarray}}
\def \be{\begin{align}}
\def \ee{\end{align}}
\def \bea{\begin{eqnarray}}
\def \eea{\end{eqnarray}}

\begin{document}

\title{Exact analysis of the interplay of charge order and unconventional pairings in the 2D Hatsugai-Kohmoto model}

\author{Carlos Eduardo S. P. Corsino}
\affiliation{Instituto de F{\'i}sica, Universidade Federal de Goi{\'a}s, 74.690-900,
Goi{\^a}nia-GO, Brazil}

\author{Hermann Freire}
\affiliation{Instituto de F{\'i}sica, Universidade Federal de Goi{\'a}s, 74.690-900,
Goi{\^a}nia-GO, Brazil}

\begin{abstract}
We provide here a study of some competing ordering tendencies exhibited by the exactly solvable 2D Hatsugai–Kohmoto (HK) model on a square lattice. To this end, we investigate the interplay between superconductivity, charge-density wave (CDW) and pair-density wave (PDW) orders as a function of interaction, doping parameter, magnetic field, and uniaxial strain. As a result, we confirm the intertwined nature of CDW and PDW fluctuating orders for intermediate-to-strong couplings. We also verify that, while an applied magnetic field favors the formation of a CDW and allows the subsequent emergence of a PDW as a secondary order, strain effects favor unidirectional PDW as a primary order over the subdominant appearance of a stripe-like CDW. These results underscore the value of the HK model as an interesting platform in order to investigate (via an exactly solvable framework) the emergence of charge order and unconventional superconductivity in fermionic systems with strong interactions. Finally, we briefly discuss an orbital generalization of the HK model, which has been recently argued to be relevant to describe the properties of realistic strongly correlated systems.

\end{abstract}

\maketitle

\section{Introduction} 

The mechanism of high-temperature superconductivity in the cuprates is probably one of the most profound enigmas in condensed matter physics \cite{keimer_review} and remains largely uncomprehended to this date. In this respect, it is fair to say that there is a reasonably broad consensus in the community that a minimal model that potentially describes the physics of these materials microscopically is the 2D Hubbard model \cite{PALee_review,anderson1987resonating} on a square lattice with the on-site interaction $U$ being of the order of the bandwidth $W$. On the other hand, this intermediate interaction regime of the model is widely acknowledged to be one of the hardest problems to be addressed theoretically (both analytically and numerically) \cite{leblanc2015solutions, arovas2022hubbard} and, for this reason, simpler models that are easier to solve and can potentially capture some important aspects of the physics of these compounds are extremely interesting in the field. One such model is the so-called 2D Hatsugai-Kohmoto (HK) model, proposed in the early 1990s \cite{hatsugai1992exactly} (a similar model was proposed in Ref. \cite{G_Baskaran}), which is an exactly solvable interacting fermionic model that describes both a Mott insulating phase at half-filling and a non-Fermi liquid (NFL) metallic phase that emerges upon doping \cite{continentino1994scaling, vitoriano2000metal, yeo2019local}. It was later demonstrated that this NFL metallic phase also turns out to be unstable towards the formation of superconductivity at low temperatures \cite{Phillips2020, li2022two}, which is qualitatively consistent with the behavior observed, e.g., in the cuprate compounds \cite{PALee_review, fradkin2015colloquium, timusk1999pseudogap}, among other systems. For this reason, our main aim in this paper will be to continue investigating some additional properties of this model in an attempt to potentially gain some insight into the remarkable physics of such strongly correlated materials.

An important aspect that we will be especially concerned with here will be the interplay of some competing electronic phases that appear in the 2D HK model at low temperatures from an exact point of view. As will become clear shortly, we will obtain that in addition to the Mott insulating and NFL phases \cite{hatsugai1992exactly,Phillips2020, li2022two}, this model can describe a variety of other ordering tendencies such as, e.g., unidirectional and bidirectional charge-density waves (CDW), spin-singlet and spin-triplet superconductivity (SC), and also pair-density-wave (PDW\footnote{The PDW phases are defined as unconventional pairing states, which possess a finite Cooper-pair center-of-mass momentum.}) phases, which demonstrate the richness of the underlying physics already captured by this simple effective model. However, we point out that not all these emergent phases will necessarily develop long-range order at low temperatures in the model\footnote{Although some electronic phases exhibit strong fluctuations in the 2D HK model, they will remain nevertheless short-ranged.}, but the main message that we want to convey here is that we will be able to identify the intertwined nature of some ordering tendencies and also obtain a hierarchy between those phases that may give some insight of what might happen in more realistic microscopic descriptions of some strongly correlated superconductors.  

Furthermore, we will study, e.g., the effects of uniaxial strain and applied magnetic field in the 2D HK model as a means to enhance (or suppress) specific phases that appear in the model in order to disentangle the different ordering tendencies that turn out to be very close in energy in the corresponding phase diagram. Altering the non-interacting band structure will also be explored here in order to understand how robust are some emergent phases described here with respect to these small modifications in the model. Finally, we will discuss a possible generalization of the HK model, which might eventually be of relevance in order to compare its low-temperature physics with more realistic strongly correlated systems, such as, e.g., the 2D Hubbard model.

Therefore, this paper will be organized as follows.
In Sec. \ref{HKmodel}, we will define the 2D HK Hamiltonian. In this part, we will be very concise and will refrain from discussing all previous results obtained for this model available in the literature (for this reason, we also refer here to some previous papers that discuss several aspects of its properties - see, e.g., \cite{hatsugai1992exactly, Phillips2020, li2022two,fermiarcs,zhu2021effects,froldi2024strong, Zhao_review_2025}). {We point out that our present paper represents a significant extension of our earlier work \cite{froldi2024strong} in that we now investigate charge order with a modulation $\mathbf{q}\neq 0$ associated with $s$- and $d$-wave symmetries and also pairing orders (i.e., both SC and PDW) associated with $p$- and $d$-wave symmetries and, in addition, we provide a comprehensive analysis of several variations of the HK model such as including an applied external magnetic field, uniaxial strain effects and modifications of the band structure}. In Sec. \ref{methodology}, we will discuss the methodology that will later be used to determine the corresponding phase diagrams of the model. In Sec. \ref{results_disc}, we will present the results {obtained within our present approach}. In Sec. \ref{Conclusions}, we end with a summary of our work and also provide an outlook on possible future directions. Finally, in Appendix \ref{Appendix_A}, we provide a demonstration of the susceptibility of the CDW phases that emerge in the 2D HK model, while, in Appendix \ref{Appendix_B}, we briefly discuss an orbital generalization of the HK model.

\section{HK Model}
\label{HKmodel}

Our starting point is the 2D HK Hamiltonian \cite{hatsugai1992exactly,Phillips2020} (for a recent review about this model, see also Ref. \cite{Zhao_review_2025}) defined on a square lattice with an additional Zeeman term, which is given by:
\begin{equation}\label{hamilto1}
\begin{aligned}
H_{HK} &= \sum_{\mathbf{k}, \sigma} \left( \xi_{\mathbf{k}} - \sigma B \right) c_{\mathbf{k}\sigma}^\dagger c_{\mathbf{k}\sigma} + U \sum_{\mathbf{k}} n_{\mathbf{k} \uparrow} n_{\mathbf{k} \downarrow}, 
\end{aligned}
\end{equation}
where $c^{\dagger}_{\mathbf{k} \sigma}$ and $c_{\mathbf{k} \sigma}$ are, respectively, the fermionic creation and annihilation operators for electrons with momentum $\mathbf{k}$ and spin projection $\sigma = \uparrow, \downarrow$. The term $\xi_{\mathbf{k}} = -2(t_x \cos k_x + t_y \cos k_y) - \mu$ describes the single-particle band dispersion, with $t_x = t - \delta$ and $t_y = t + \delta$ denoting the hopping amplitudes along the $x$- and $y$-directions, respectively. We point out that these hopping amplitudes are allowed to be modified by the most direct effect of the strain $\delta$ (which is the increase in the orbital overlap integral by compression, generating anisotropy in the hopping in $x$ and $y$ directions \cite{Scalettar_strain}) and $\mu$ is the chemical potential. The term proportional to $B$ corresponds to the Zeeman term due to an external magnetic field. The number operator is defined as $n_{\mathbf{k} \sigma} = c^{\dagger}_{\mathbf{k} \sigma} c_{\mathbf{k} \sigma}$, and $U > 0$ denotes the local repulsive interaction in momentum space. All external perturbations (magnetic field, strain, etc.) included in the 2D HK model preserve the exact solvability of the model. This enables us to investigate these modifications of such a strongly correlated model while maintaining analytical tractability. In order to analyze the emergence of some correlated phases in this model, we will add to Eq. \eqref{hamilto1} a coupling in the charge-density wave and pairing channels encapsulated in the general term denoted by $H_{V}=H_{SC}+H_{CDW}$, whose terms in the r.h.s. of the equation will be explained below. Here, we will focus only on the aspects of the physics of the model that will be our main concern in this work, since
other properties and additional technical details can also be found in several papers available in the literature (see, e.g., Refs. \cite{hatsugai1992exactly, Phillips2020, li2022two,fermiarcs,  zhu2021effects,Zhao_review_2025}) and also in a previous paper by the authors \cite{froldi2024strong}. 

There are various possible microscopic mechanisms for Cooper pairing (phonon-mediated, spin-fluctuation-mediated, etc.). In what follows, we will be agnostic with regard to the origin of pairing and will simply assume an attractive BCS-type interaction \cite{cooperoriginal, bcs1957}, which is written in momentum space as 
\begin{equation}
H_{SC} = -V_p\sum_{{\mathbf{k}, \mathbf{k}',\sigma,\sigma'}}  \gamma_{\mathbf{k}}\gamma_{\mathbf{k'}}\,c^{\dagger}_{\mathbf{k}+\mathbf{q}/2,\sigma}
 c^{\dagger}_{\mathbf{-k}+\mathbf{q}/2,\sigma'}
 c_{\mathbf{-k'}+\mathbf{q}/2,\sigma'}c_{\mathbf{k'}+\mathbf{q}/2,\sigma},
\end{equation}
where $V_p>0$ (we point out that we also consider here another simplifying assumption that the pairing interaction for both spin-singlet and spin-triplet channels is equal). For spin-singlet pairing, the pair operator associated with momentum $\mathbf{q}$ -- which can be zero for a uniform SC phase or finite for
a PDW phase -- is defined as $\Delta_{\mathbf{k}}(\mathbf{q}) = \gamma_{\mathbf{k}}\,c_{-\mathbf{k} + \frac{\mathbf{q}}{2}, \downarrow} \, c_{\mathbf{k} + \frac{\mathbf{q}}{2}, \uparrow}$, with $\gamma_{\mathbf{k}}$ being a form factor that encodes even-parity symmetry ($s$- and $d$-wave) \cite{scalapino1995case}. For spin-triplet pairing, the corresponding pair operator is defined as $\Delta_{\mathbf{k}}(\mathbf{q}) = \gamma_{\mathbf{k}}\,c_{-\mathbf{k} + \frac{\mathbf{q}}{2}, \uparrow} \, c_{\mathbf{k} + \frac{\mathbf{q}}{2}, \uparrow}$, with $\gamma_{\mathbf{k}}$ being an odd-parity ($p$-wave) form factor \cite{pwave1}. Therefore, we will consider the following form factors associated with each of the above-mentioned symmetries:
\begin{equation}\label{form_factors}
    \gamma_{\mathbf{k}} = 
    \begin{cases} 
        1, & \text{($s$-wave)}; \\
        \cos k_x - \cos k_y, & \text{($d$-wave)}
        ; \\
        \sin k_x + i \sin k_y, & \text{($p$-wave)}.
    \end{cases}
\end{equation}
By following a similar strategy explained before, we will also assume an attractive coupling in the CDW channel, which is given as follows:
\begin{equation}\label{DW_eq}
H_{\text{CDW}} =-V_c\sum_{\mathbf{k},\mathbf{k'},\sigma,\sigma'}  \gamma_{\mathbf{k}}\gamma_{\mathbf{k'}}\, {c}^\dagger_{\mathbf{k} + {\mathbf{q}/2}, \sigma} \, {c}^\dagger_{\mathbf{k'}  -\mathbf{q}/2, \sigma'} \, {c}_{\mathbf{k'} + {\mathbf{q}/2}, \sigma'}\,{c}_{\mathbf{k} -\mathbf{q}/2,\sigma} ,
\end{equation}
where $V_c>0$ (for a fixed $\mathbf{q}\neq 0$), and the CDW operator associated with a modulation $\mathbf{q}$ is defined as $\rho_{\mathbf{k}}(\mathbf{q}) = \gamma_{\mathbf{k}}\sum_{\sigma} {c}^\dagger_{\mathbf{k} + {\mathbf{q}}/{2}, \sigma} \, {c}_{\mathbf{k} - {\mathbf{q}}/{2}, \sigma}$ [we will consider here both $s$- and $d$-wave symmetries for the CDW, whose form factors are defined in Eq. \eqref{form_factors}].

The exact Green’s function in the Hatsugai–Kohmoto model can be written in closed form as \cite{hatsugai1992exactly,Phillips2020} 
\begin{align}\label{ExactGF}
G_{\mathbf{k},\sigma}(i\omega) 
&= \frac{1 - \langle n_{\mathbf{k},\bar{\sigma}} \rangle}{i\omega - \xi_{\mathbf{k},\sigma}}
+ \frac{\langle n_{\mathbf{k},\bar{\sigma}} \rangle}{i\omega - \xi_{\mathbf{k},\sigma} - U},
\end{align}
where $\xi_{\mathbf{k},\sigma} = \xi_{\mathbf{k}} - \sigma B $ and $\langle n_{\mathbf{k},\bar{\sigma}} \rangle$ denotes the average occupation of momentum $\mathbf{k}$ with opposite spin $\bar{\sigma}=-\sigma$. This occupation is given by the expression:
\begin{equation} \label{nk}
\langle n_{\mathbf{k},\sigma} \rangle = \frac{
e^{-\beta \xi_{\mathbf{k}, \sigma}} + e^{-\beta (2\xi_{\mathbf{k}} + U)}
}{
1 + e^{-\beta \xi_{\mathbf{k}, \uparrow}} + e^{-\beta \xi_{\mathbf{k}, \downarrow}} + e^{-\beta (2\xi_{\mathbf{k}} + U)}
}.
\end{equation}
Let us analyze the above result in the absence of external perturbations in the zero-temperature limit. In this case, the average occupation is given by the following piecewise expression \cite{hatsugai1992exactly,Phillips2020, li2022two}:
\begin{align} \label{number1}
    \langle n_{\mathbf{k}, \sigma} \rangle = 
        \begin{cases}
            0, & \mu < \epsilon_\mathbf{k} < \frac{W}{2} \hspace{1.5cm}(\mathbf{k} \in \Omega_0),\\
            \frac{1}{2}, &  \mu - U < \epsilon_\mathbf{k} < \mu \hspace{0.97cm}(\mathbf{k} \in \Omega_1),\\
            1, & - \frac{W}{2} < \epsilon_\mathbf{k} < \mu - U \hspace{0.54cm}(\mathbf{k} \in \Omega_2),
        \end{cases}
\end{align}
where {$W$ is the bandwidth of the model and} the momentum regions $\Omega_0$, $\Omega_1$, and $\Omega_2$ correspond, respectively, to empty, single-particle-occupied, and double-particle-occupied momentum states. We note that due to the existence of the $\Omega_1$ region, the ground state displays a massive degeneracy, which is lifted by the external perturbation denoted by $H_V$. Finally, in order to connect the above division of the momentum space with the hole doping parameter $x$ in the 2D HK model, we obtain by demanding that the $\Omega_1$ region lies within the bandwidth the following conditions at $T = 0$ \cite{froldi2024strong}:
\begin{align} \label{doping}
x = 
\begin{cases}
    \frac{1}{2} - \tilde{\mu}, & \text{if } \tilde{\mu} - u < -\frac{1}{2} \text{ and } -\frac{1}{2} < \tilde{\mu} \leq \frac{1}{2},\\
    u - 2\tilde{\mu}, & \text{if } \tilde{\mu} - u > -\frac{1}{2} \text{ and } -\frac{1}{2} < \tilde{\mu} \leq \frac{1}{2},
\end{cases}
\end{align}
where {we defined the dimensionless quantities} $\tilde{\mu} = \mu/W$ and $u = U/W$.

\section{Methodology}
\label{methodology}

To determine the ordering tendencies in the 2D HK model, we begin by computing several order-parameter susceptibilities. The pair susceptibility for the spin-singlet case is defined as  
\begin{align}
\chi_{p,(s)}^{(0)}(i\nu_n,\mathbf{q}) 
&= \int_{0}^{\beta} d\tau\,e^{i\nu_n\tau}
\big\langle T\,\Delta_{\mathbf{q}}(\tau)\,\Delta_{\mathbf{q}}^{\dagger}(0)\big\rangle_{0}\nonumber \\[0.5ex]
&= -\frac{1}{\beta} \sum_{\omega_m} \sum_{\mathbf{k}}
\gamma_{\mathbf{k}}\gamma_{\mathbf{k}}\,G_{\mathbf{k} + \frac{\mathbf{q}}{2},\,\uparrow}(i\omega_m)\,
G_{-\mathbf{k} + \frac{\mathbf{q}}{2},\,\downarrow}(i\nu_n - i\omega_m),
\end{align}
where the form factors can be either $s$- or $d$-wave. The product of the Green’s functions yields the expression for the ``bare'' pair susceptibility:
\begin{equation} 
\chi_{p,(s)}^{(0)}(i\nu_n,\mathbf{q})
= \sum_{\mathbf{k},a,b}
\gamma_{\mathbf{k}}\gamma_{\mathbf{k}}\,n_{\mathbf{k} + \frac{\mathbf{q}}{2},\uparrow}^a \;
n_{-\mathbf{k} + \frac{\mathbf{q}}{2},\downarrow}^b \;
\frac{
f\!\bigl(\xi_{\mathbf{k} + \frac{\mathbf{q}}{2},\uparrow}^a\bigr)
+ f\!\bigl(\xi_{-\mathbf{k} + \frac{\mathbf{q}}{2},\downarrow}^b\bigr)
- 1
}{
i\nu_n 
- \xi_{\mathbf{k} + \frac{\mathbf{q}}{2}}^a
- \xi_{-\mathbf{k} + \frac{\mathbf{q}}{2}}^b
},
\end{equation}
where $a$ and $b$ label the upper (denoted by $u$) and lower (denoted by $l$) bands, defined by $\xi_{\mathbf{k}}^u = \xi_{\mathbf{k}} + U$ and $\xi_{\mathbf{k}}^l = \xi_{\mathbf{k}}$.
Note that the energies in the denominator do not carry spin indices. This is because the particles being paired have opposite spins. As a result, the magnetic contributions enter with opposite signs and cancel each other out.

For the spin-triplet case, the ``bare'' pair susceptibility becomes
\begin{equation} 
\chi_{p,(t)}^{(0)}(i\nu_n, \mathbf{q})
= \sum_{\mathbf{k}, a, b}
\gamma_{\mathbf{k}}\gamma_{\mathbf{k}}\,n_{\mathbf{k} + \frac{\mathbf{q}}{2},\downarrow}^a \;
n_{-\mathbf{k} + \frac{\mathbf{q}}{2},\downarrow}^b \;
\frac{
f\!\bigl(\xi_{\mathbf{k} + \frac{\mathbf{q}}{2},\uparrow}^a\bigr)
+ f\!\bigl(\xi_{-\mathbf{k} + \frac{\mathbf{q}}{2},\uparrow}^b\bigr)
- 1
}{
i\nu_n 
- \xi_{\mathbf{k} + \frac{\mathbf{q}}{2},\uparrow}^a
- \xi_{-\mathbf{k} + \frac{\mathbf{q}}{2},\uparrow}^b
},
\end{equation}
where now the form factor is $p$-wave.

The ``bare'' charge susceptibility is obtained from the density–density correlation function:
\begin{align}
\chi_{c}^{(0)}(i\nu_n,\mathbf{q})
&= \int_{0}^{\beta} d\tau\,e^{i\nu_n\tau}
\big\langle T\,\rho_c(\mathbf{q},\tau)\,\rho_c(-\mathbf{q},0)\big\rangle_{0} \nonumber\\[0.5ex]
&= -\frac{1}{\beta}\sum_{i\omega_m}\sum_{\mathbf{k},\sigma}\gamma_{\mathbf{k}}\gamma_{\mathbf{k}}\,
G_{\mathbf{k} - \frac{\mathbf{q}}{2},\,\sigma}(i\omega_m)\;
G_{\mathbf{k} + \frac{\mathbf{q}}{2},\,\sigma}(i\omega_m + i\nu_n),
\end{align}
which leads to the following expression for the charge susceptibility:
\begin{align}
\chi_c^{(0)}(i\nu_n, \mathbf{q}) 
&= - \sum_{\mathbf{k},\sigma} \sum_{a,b}\gamma_{\mathbf{k}}\gamma_{\mathbf{k}}\, 
n_{\mathbf{k} - \frac{\mathbf{q}}{2},\bar{\sigma}}^a \, 
n_{\mathbf{k} + \frac{\mathbf{q}}{2},\bar{\sigma}}^b
\frac{
f\bigl(\xi_{\mathbf{k} - \frac{\mathbf{q}}{2},\sigma}^a\bigr) - 
f\bigl(\xi_{\mathbf{k} + \frac{\mathbf{q}}{2},\sigma}^b\bigr)
}{
i\nu_n + 
\xi_{\mathbf{k} - \frac{\mathbf{q}}{2}}^a - 
\xi_{\mathbf{k} + \frac{\mathbf{q}}{2}}^b
}.
\end{align}
Note that once again the energies in the denominator do not include spin indices. In this case, although the Green’s functions are spin dependent, the energy dispersions appear with opposite signs for the same spin projection, and the spin-dependent contributions cancel out.

It is important to note that we have computed only the ``{bare}'' susceptibility in each case. As shown in Ref.~\cite{Phillips2020}, the full susceptibility $\chi_p(i\nu_n,\mathbf{q})$ is related to the bare one via a Dyson-like equation:
\begin{equation}
\chi_p(i\nu_n,\mathbf{q}) =  \frac{\chi_p^{(0)}(i\nu_n,\mathbf{q})}{1 - V_p\, \chi_p^{(0)}(i\nu_n,\mathbf{q})}.
\end{equation}
A similar result also holds for the CDW susceptibilities, as shown in our derivation in Appendix \ref{Appendix_A}. Therefore:
\begin{equation}
\chi_c(i\nu_n,\mathbf{q}) =  \frac{\chi_c^{(0)}(i\nu_n,\mathbf{q})}{1 - V_c\, \chi_c^{(0)}(i\nu_n,\mathbf{q})}.
\end{equation}
An immediate consequence of these relations is that the critical temperature associated with the instability of each phase is determined by the following condition:
\begin{equation}
    \chi_i^{(0)}(i\nu_n = 0,\mathbf{q})\Big|_{T = T_c^{(i)}} = \frac{1}{V_i},
\end{equation}
where the index $i$ refers to the type of the order-parameter susceptibility [e.g., pair (denoted by ``$p$'') and CDW (denoted by ``$c$'')] and the corresponding critical temperature $T_c^{(i)}$. To keep the parameter space of the present model manageable, we will also assume from now on that $V_p=V_c=V$, for simplicity.

Finally, another important quantity to consider is the ferromagnetic susceptibility. As shown in Ref.~\cite{guerciHK}, in the calculation of both the uniform charge and spin susceptibilities (i.e., for $\mathbf{q} = 0$), the linear response theory applied to the HK model breaks down. However, this problem can be circumvented by employing the following thermodynamic expression for the uniform spin susceptibility, which is given by:
\begin{equation}
    \chi_{s} = \lim_{B\rightarrow 0}\frac{\partial M}{\partial B} = \lim_{B\rightarrow 0} \sum_{\mathbf{k}} \left( \frac{\partial n_{\mathbf{k}\uparrow}}{\partial B} - \frac{\partial n_{\mathbf{k}\downarrow}}{\partial B} \right)\mu_B,
\end{equation}
where $M$ is the magnetization, $\mu_B$ is the Bohr magneton, and $n_{\mathbf{k}\sigma}$ is defined in Eq.~\eqref{nk}.

\begin{figure}[b]
\centering
\centering \includegraphics[width=0.35\linewidth]{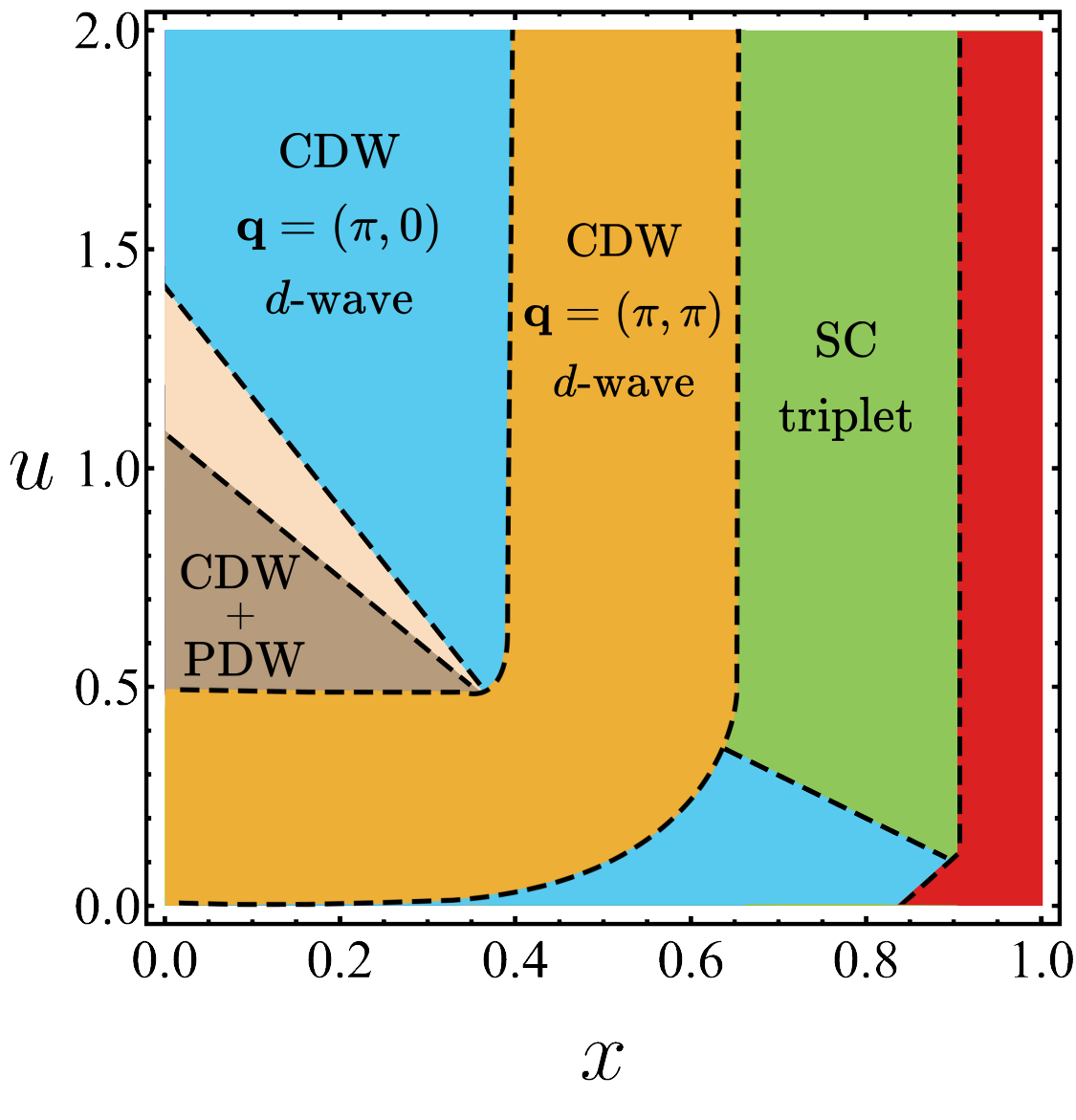}
\caption{
  Phase diagram {of dimensionless interaction $u=U/W$ versus doping parameter $x$} of the competing ordering tendencies that emerge in the 2D HK model with $V = 0.65$, with no magnetic field ($B=0$) and no strain ($\delta=0$). The label `CDW + PDW' means that both CDW and PDW at $\mathbf{q}=(\pi,\pi)$ with $s$-wave symmetry are found within this region. The beige color denotes a region where a CDW at $\mathbf{q}=(\pi,\pi)$ with $s$-wave symmetry (with a subdominant PDW) appears in the model. The red part denotes $s$-wave singlet SC. At intermediate-to-strong coupling, CDW orders allow the emergence of PDW as a secondary
order with the same symmetry and the same wavevector. The dashed lines are only a guide to the eye.
}
\label{no_magn}
\end{figure}

\section{Results and Discussion}
\label{results_disc}

To begin with, we discuss how the competition among the different ordering tendencies in the 2D HK model takes place in the absence of strain ($\delta=0$) and external magnetic field ($B=0$). As will become clear soon, from the analysis of the order-parameter susceptibilities (see, e.g., Refs. \cite{Freire_Ferraz_2008,Correa_Freire_Ferraz_2008,Freire_Ferraz_2005,Freire_Caetano_2019} in the context of other 2D correlated models), we obtain strong evidence of multiple short-order tendencies of comparable strengths within the present model. For this reason, we will focus henceforth on the dominant orders only, for each regime of the dimensionless interaction $u$ and doping parameter $x$. 

For infinitesimal $V\rightarrow 0$, we confirm that the only instability of NFL phase in the model away from half-filling is towards the formation of superconductivity, which agrees with previous results regarding this model that are available in the literature \cite{Phillips2020, li2022two, zhu2021effects,froldi2024strong}. However, for a finite $V$, we obtain a variety of additional short-range orders that emerge in the corresponding phase diagram, which are summarized in our Fig.~\ref{no_magn}. We observe that at weak coupling near half-filling and beyond, a bidirectional fluctuating $d$-wave CDW order appears near $\mathbf{q}\approx (\pi,\pi)$. At weak-coupling and near half-filling, the essential driver for this order is the presence of approximate nesting of the Fermi surface. 

As the local interaction $u$ becomes intermediate-to-strong, an intertwined CDW with a PDW near $\mathbf{q}\approx (\pi,\pi)$ dominates and this behavior takes place from a small doping up to a critical doping parameter $x_c$, in agreement with our previous results in  Ref.~\cite{froldi2024strong}. This is labeled as ‘CDW + PDW’ in Fig.~\ref{no_magn} and both phases possess $s$-wave symmetry. 
This feature of the PDW being intertwined with the CDW region displayed by the 2D HK model is very interesting, since it indicates a close connection between these two phases \cite{Agterberg_2020,de_Carvalho_2014,Freire_2015,Freire_loop_current_2015,Freire_loop_current_2016,Kloss_2016}. Therefore, the 2D HK model provides good evidence supporting this interdependent nature of CDW and PDW phases within the underdoped regime and intermediate $u$. A similar mechanism was proposed to describe some aspects of the physics of the pseudogap regime in the cuprates in the underdoped regime (see, e.g., Refs. \cite{Agterberg_2020,de_Carvalho_2014,Freire_2015,Freire_loop_current_2015,Freire_loop_current_2016,Kloss_2016}).

Moreover, for an even higher interaction $u$, another charge ordered phase emerges in the model:  a unidirectional (stripe-like) CDW near $\mathbf{q}=(\pi,0)$ with $d$-wave symmetry (with an accompanying subdominant PDW order with the same symmetry and associated with the same wavevector). For such strong couplings, the CDW order is not necessarily responsive to details of the Fermi surface structure, and depends to a great extent only on the strength of $V$. Lastly, far from half-filling, first a $p$-wave spin-triplet SC phase emerges and then, with further doping, an $s$-wave spin-singlet SC appears.
The former SC phase is related to the presence of ferromagnetic fluctuations in the model, which mediate the pairing interaction (cf. the discussion about the ferromagnetic fluctuations for the 2D HK model in the case of $n=1$ orbital in Appendix \ref{Appendix_B}). As soon as these fluctuations become extremely weak as a result of strong overdoping, a conventional SC phase (i.e., $s$-wave) emerges at low temperatures.

\begin{figure}[b]
\centering
\centering \includegraphics[width=0.35\linewidth]{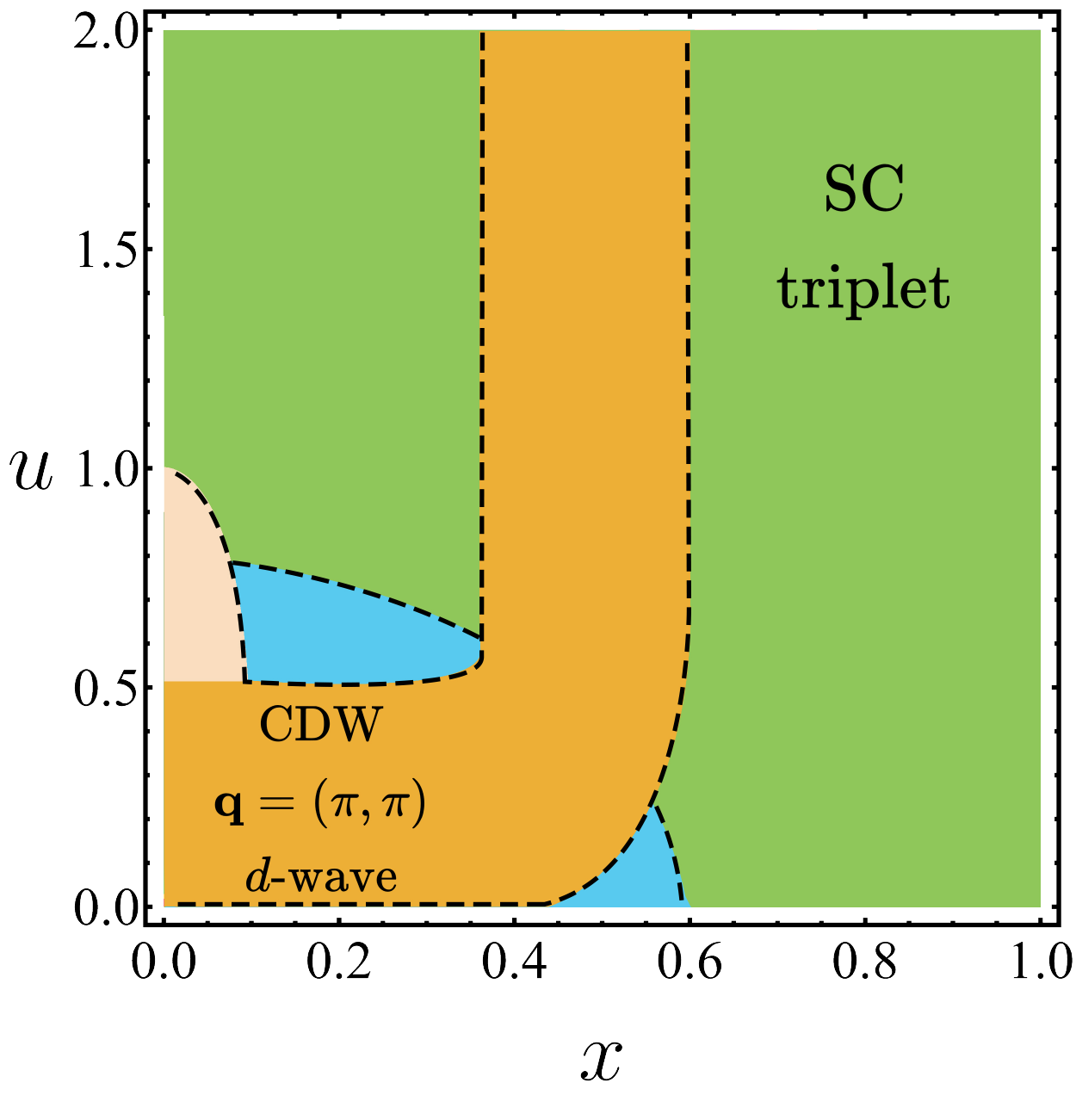}
\caption{
  Phase diagram {of dimensionless interaction $u=U/W$ versus doping parameter $x$} of the competing orders in the 2D HK model for $B = 0.05$, $V = 0.4$ and $\delta=0$. The light blue regions correspond to $d$-wave CDW with $\mathbf{q} = (\pi, 0)$, and the beige color denotes the $s$-wave CDW with $\mathbf{q} = (\pi, \pi)$.  The dashed lines are only a guide to the eye.}
\label{magnetic_1}
\end{figure}

\subsection{Magnetic field effect}

The next step is to analyze the effect of a finite external magnetic field applied to the model ($B\neq 0$). Overall, we observe that both the spin-singlet SC and PDW phases are suppressed by an external magnetic field (see Fig. \ref{magnetic_1}). For the spin-triplet SC and CDW cases, the situation becomes more interesting, since these phases strongly compete with each other, depending on the interaction $u$ and the doping parameter $x$. For very small interaction potentials ($V \to 0$), the spin-triplet SC phase becomes the dominant instability of the model throughout the phase diagram. This is related to the presence of ferromagnetic fluctuations in the model, as was already discussed in the previous section, which become even stronger as a result of the magnetic field.
However, for slightly larger $V$, the CDW phase becomes increasingly relevant and starts to dominate the $p$-wave SC phase for some regimes, as shown in Fig. \ref{magnetic_1}. 

It is important to note that this result holds for small magnetic fields. As the strength of $B$ increases, a stronger interaction $V$ is required for the CDW phase to occupy a significant portion of the phase diagram, otherwise the spin-triplet SC phase prevails.
Moreover, the $p$-wave SC phase is now clearly enhanced near the Mott-insulating regime, where this pairing state becomes the dominant order. The CDW fluctuations also exhibit a subtler response to the application of a finite $B$: although it is modestly enhanced at intermediate dopings, the main effect of the magnetic field is, in fact, to shift the regions where this phase is already favored due to an approximate nesting condition. The only exception to this rule is the $d$-wave unidirectional (i.e., stripe-like) CDW at the wave vector $\mathbf{q}=(\pi,0)$, which now appears in a region of the phase diagram of weak-to-intermediate $u$ and at moderate-to-large doping $x$.

\begin{figure}[t]
\centering
\centering \includegraphics[width=0.35\linewidth]{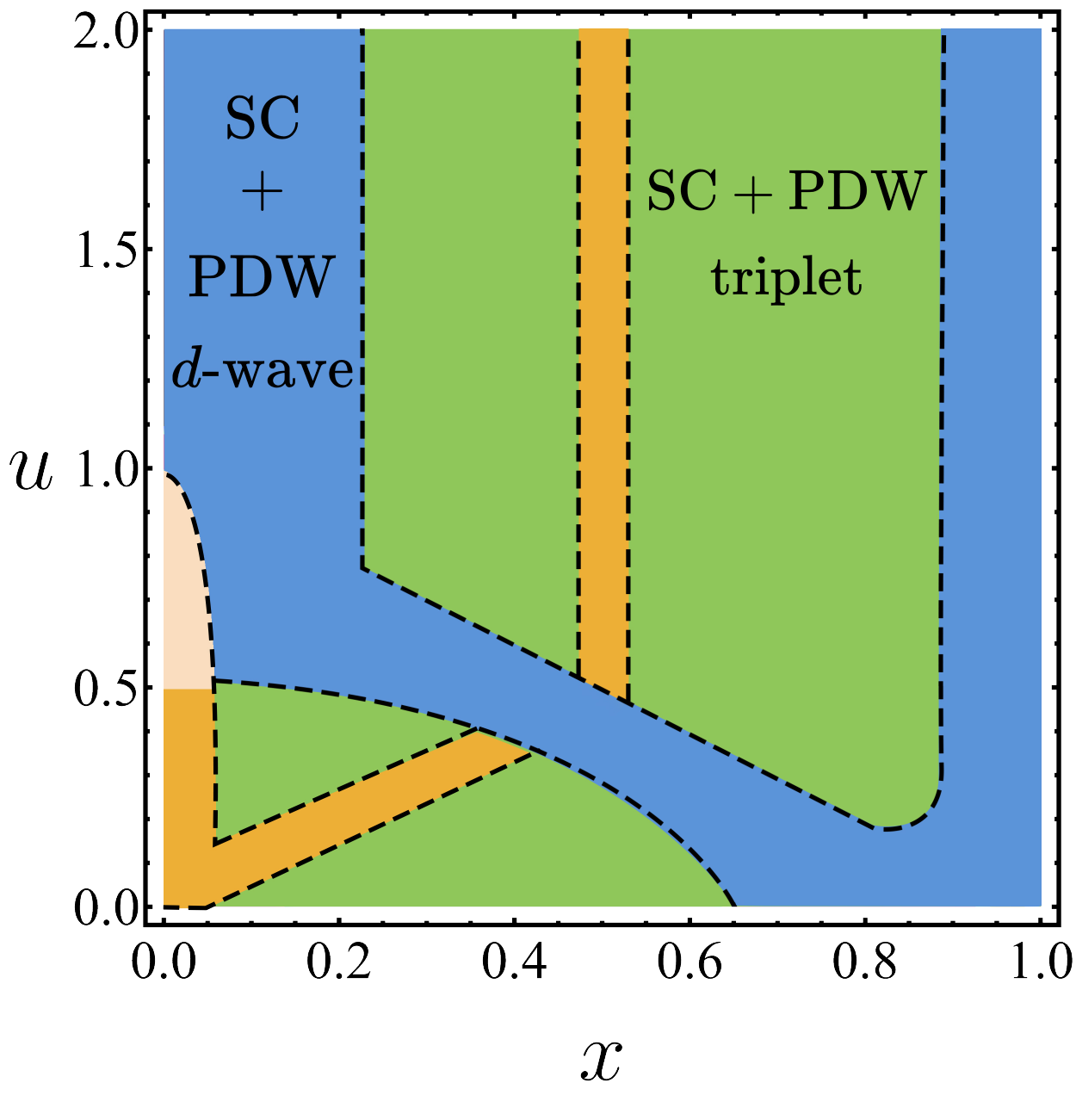}
\caption{Phase diagram {of dimensionless interaction $u=U/W$ versus doping parameter $x$} of the 2D HK model for the strain anistropy given by $\delta_c=W/8$ and interaction $V=0.15$. Both PDW orders represented above are unidirectional and have ordering vector near $\mathbf{q} = (\pi, 0)$. The yellow region denotes the $d$-wave CDW phase near $\textbf{q} = (\pi, \pi)$, as the beige color corresponds the $s$-wave CDW near $\mathbf{q} = (\pi, \pi)$.  The dashed lines are only a guide to the eye.}\label{strain_1}
\end{figure}

\subsection{Strain effect}

In the previous section, we have obtained that the SC correlations appear stronger whenever the CDW correlations are weaker and vice-versa (i.e., they both compete with each other). Since the application of strain $\delta$ is well known to reduce charge ordering tendencies, we now turn to a discussion of these effects in the present model.
We show the corresponding results in Fig.~\ref{strain_1}. From this plot, we observe that the spin-triplet SC phase at intermediate-to-large doping remains to some extent robust throughout the phase diagram. Despite this, we note that the $d$-wave spin-singlet SC phase now appears in some regions, and, in addition, it becomes displaced toward the edges of the phase diagram as the strain anisotropy increases. For even larger $\delta$, the $d$-wave SC region continues this trend, and eventually, for higher values ($\delta > 0.15$), it disappears completely.

The CDW phase follows the aforementioned expected behavior. Although some regions (where CDW dominates) remain qualitatively similar as before, it now becomes clearly more fragile, disappearing almost entirely for smaller $V$. Its persistence along some characteristic parts of the phase diagram arises because the CDW with $\mathbf{q} = (\pi, \pi)$ is intrinsically a bit more robust in these regions; thus, even when the application of strain tends to suppress it, some remnants of the phase are maintained. Consequently, if the interaction $V$ is fixed and $\delta$ is increased further, the CDW phase gradually vanishes altogether from the phase diagram.

It is also interesting to point out that a unidirectional PDW phase associated with the ordering vector $\mathbf{q} = (\pi, 0)$ is strongly favored under strain application for large $u$ and within the underdoped regime. In fact, when the strain increases to a critical $\delta_c = W/8$, a particularly interesting phenomenon emerges: the PDW phase becomes exactly degenerate with the SC phase and spans almost the whole phase diagram. In Fig.~\ref{strain_1}, we denote this fact by the label `SC + PDW'.

Now, focusing specifically on the interconnection between the spin-singlet PDW phase and the CDW, the hierarchy between these two phases becomes notably different from the previously discussed cases. 
The application of strain now changes the balance 
between those intertwined phases, with PDW now becoming the primary order for stronger interactions with the subsequent appearance of a stripe-like
CDW order as the secondary phase. Nevertheless, both phases remain associated with the same symmetry and possess the same modulation described by the wavevector $\mathbf{q}$. The exception to this rule is the regime of weak-to-intermediate interaction
and intermediate doping, where the previous behavior remains true for the CDW with $\mathbf{q} = (\pi, \pi)$ wavevector, since such a bidirectional CDW is still stabilized in regions of the phase diagram with approximate nesting of the underlying Fermi surface of the 2D HK model.

{From an experimental point of view, it is still very challenging to determine whether unidirectional PDW can be favored by strain effects in the context of the cuprate superconductors. The experimental measurements of a PDW state in these compounds have been limited so far to surface-sensitive scanning tunneling microscopy (STM) techniques (see, e.g., Ref. \cite{Hamidian_2016}). Nevertheless, we point out that nuclear magnetic resonance (NMR) experiments are a promising pathway in order to demonstrate the PDW in the bulk of the cuprate compounds under strain and these measurements are currently underway. Therefore, we believe that the results of the HK model could potentially provide a guide for future research into the PDW phase in these systems.}

\begin{figure}[t]
\centering
\centering \includegraphics[width=0.35\linewidth]{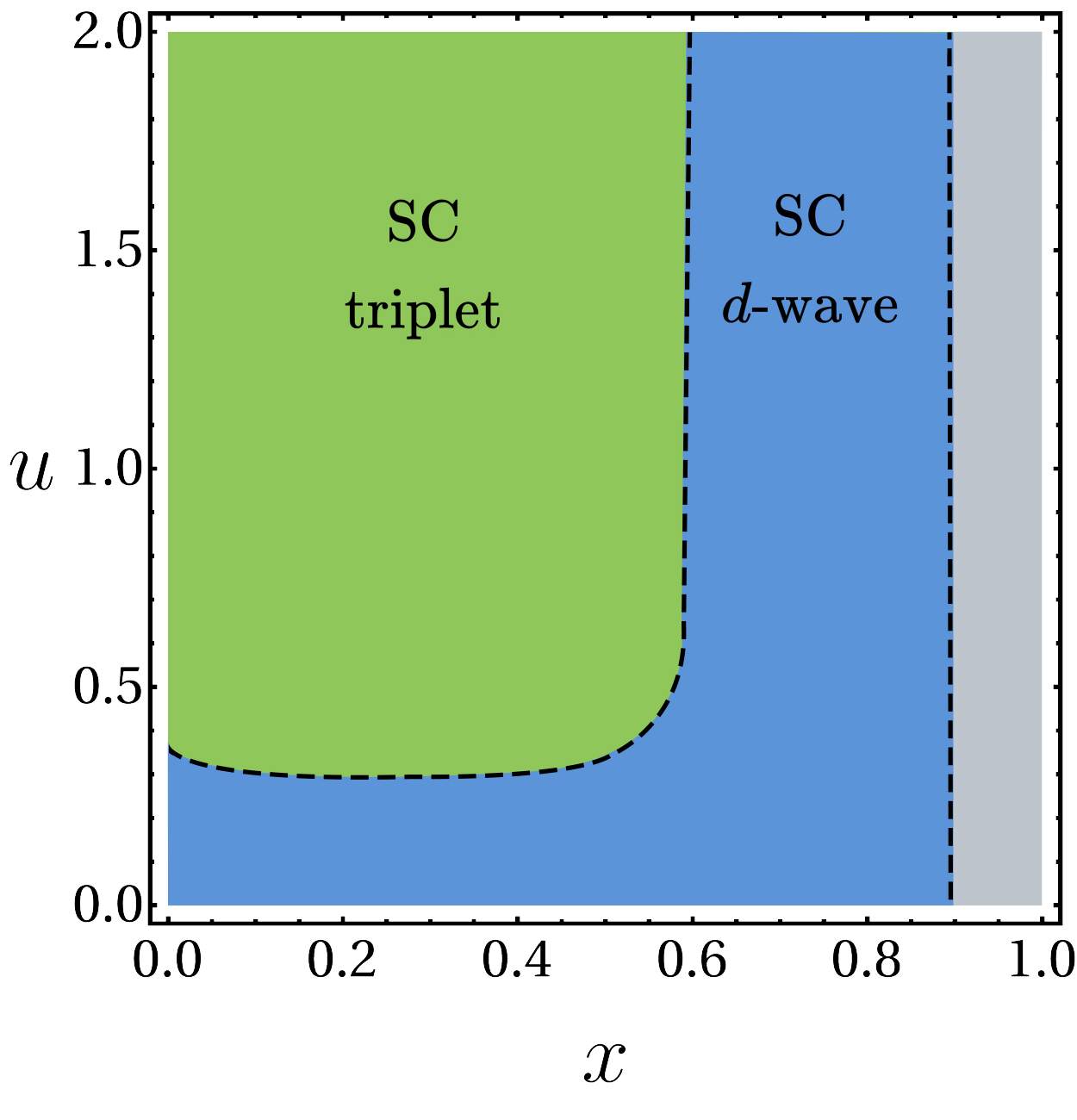}
\caption{Phase diagram {of dimensionless interaction $u=U/W$ versus doping parameter $x$} of the 2D HK model for $t'=0.45t$ and $V=0.15$ (with $\delta=0$ and $B=0$). The grey region corresponds to the NFL metallic phase.  The dashed lines are only a guide to the eye.}\label{next_1}
\end{figure}

\subsection{Inclusion of next-nearest neighbor hopping $t'$}

Another interesting case that will be considered here refers to the inclusion of next-nearest-neighbor hopping $t'$ in the band structure in the model (assuming, for simplicity, no uniaxial strain and no applied magnetic field). In this case, the energy dispersion of the 2D HK takes the form  
\begin{equation}
\xi_{\mathbf{k}} = -2t(\cos k_x + \cos k_y) + 4t' \cos k_x \cos k_y - \mu.
\end{equation}
Our results obtained for this case are shown in Fig. \ref{next_1}. From this plot, we observe that increasing $t'$ 
significantly weakens CDW correlations throughout the phase diagram. Indeed, as emphasized previously, since CDW ordering tendencies compete with pairing correlations with momentum $\mathbf{q}=0$ in the model, SC becomes stronger in this case. We note that for intermediate-to-strong interaction $u$, spin-triplet SC emerges upon doping the Mott insulator away from half-filling up to critical doping. This is related to the strong ferromagnetic fluctuations that exist in the model within this regime already mentioned before, which mediate the Cooper pairing in this case. This is expected to be an artifact of the 2D HK model, which can be corrected by generalizing the model and making it more realistic. Recently, several works in the literature have appeared \cite{bradlyn2023, hackner2025solving, DynamicalPP2025, realspace2024} that put forward a generalization of the 2D HK model via an introduction of a large number $n$ of orbitals, where it has been suggested that such an orbital HK (OHK) model, with some assumptions and in a specific limit, would approach the behavior, e.g., of the paradigmatic 2D Hubbard model defined with $N/n$  clusters ($N$ being the total number of sites) \cite{mai2024newapproachstrongcorr}. In agreement with the latter work, we show in Appendix \ref{Appendix_B} by reproducing successfully the expected behavior for such systems that, for strong interaction (i.e., $u>1$), ferromagnetic fluctuations turn out to be indeed suppressed in the model for $n\geq 4$. This further supports the hypothesis that, for a more realistic OHK model, spin-triplet SC order is not expected to emerge within this regime for a finite $U$. Lastly, we point out that a $d$-wave SC phase emerges in the 2D HK model for large $t'$ with further doping or for small $u$.

\section{Summary and Outlook}\label{Conclusions}

In this work, we have examined the interplay between several competing ordering tendencies in the 2D HK model (that displays a Mott insulating phase at half-filling and an NFL metallic phase upon doping) under the influence of some external perturbations, including magnetic fields, uniaxial strain, and next-nearest-neighbor hopping. By means of an exact analytical calculation of the corresponding order-parameter susceptibilities, we have constructed detailed phase diagrams that capture the emergence and competition of SC, CDW, and PDW orders in the model. {However, we point out that the 2D HK model has some limitations compared to more realistic strongly correlated models (such as, e.g., the 2D Hubbard model) in view of the presence of the local-in-momentum coupling $U$, whereas in fundamental models to describe, e.g., the cuprate compounds, a strong momentum dependence in the interaction is necessary.}

A central result of our analysis is that such a relatively simple model can reproduce some key physical features, potentially associated with more complex strongly correlated systems. These include the competition between SC and CDW orders, the emergence of an intertwined connection between PDW and CDW orders at intermediate-to-strong coupling in the underdoped regime, and even a correspondence with the physics of the 2D Hubbard model by means of a generalization of the 2D HK model by including a large number of orbitals. This is both remarkable and significant: the exact solvability of the {orbital} HK model may provide an analytical scheme to investigate the physics of the 2D Hubbard model at intermediate-to-large couplings without the need for heavy numerical simulations \cite{mai2024newapproachstrongcorr}. 

Therefore, the HK model offers a uniquely transparent and tractable platform for exploring fluctuation-induced strongly correlated superconductivity, and also its interplay with both CDW and PDW orders. Future directions include investigating the correlated phases analyzed here by incorporating additional orbital degrees of freedom to the HK model and also extending it to include spin–orbit-coupling (see, e.g., Refs. \cite{Mai_top_2023,Top_Mai_Haldane,jablonowski2023topological, spinorbit2023}) potentially bridging the gap between the HK model and real materials that exhibit strong correlations, topological properties, and unconventional pairing phases. Another interesting direction of research refers to the calculation of the transport properties associated with the NFL phase displayed by the 2D HK model and also its possible generalizations (see, e.g., Refs. \cite{ips_hermann1,Freire_PLA_2021,Hermann_review} for an analysis of such properties in the context of another NFL phase).


\section*{Acknowledgments}  

H.F. acknowledges funding from the Conselho Nacional de Desenvolvimento Cient\'{i}fico e Tecnol\'{o}gico (CNPq) under the grants: No. 311428/2021-5, No. 404274/2023-4, and No. 305575/2025-2.

\appendix
\section{Derivation for the  CDW susceptibilities at low temperatures}\label{Appendix_A}

Let us now derive exactly the equation for the charge susceptibility $\chi_c(i\nu_n)$ at $T\ll U$, $W$ in the presence of a CDW interaction (see Eq. \eqref{DW_eq}). 
Our goal is to evaluate the Matsubara charge susceptibility at finite coupling, i.e.
\begin{equation}
    \chi_c = \int_0^\beta d\tau \, e^{i\nu_n \tau} \langle \hat{T}_\tau \rho_c(\tau, \mathbf{q}) \rho_c(0, -\mathbf{q}) \rangle_{V_c},
\end{equation}
and relate it to the bare susceptibility $\chi_c^{(0)}(i\nu_n)$, defined by the same expression evaluated at zero interaction (i.e., $V_c= 0$).

We proceed by expanding the thermal expectation value in the interaction picture. The time-ordered correlator is given by
\begin{equation}
    \langle \hat{T}_\tau \rho_c(\mathbf{q}, \tau) \rho_c(-\mathbf{q}, 0) \rangle = \frac{\langle \hat{T}_\tau S(\beta) \rho_c(\mathbf{q},\tau) \rho_c(-\mathbf{q},0) \rangle_0}{\langle \hat{T}_\tau S(\beta) \rangle_0},
\end{equation}
with the subscript $0$ denoting expectation values taken with respect to the non-interacting Hamiltonian and the $S$-matrix is defined as
\begin{equation}
    S(\beta) = \hat{T}_\tau \exp\left( -\int_0^\beta d\tau' \, H_{\text{CDW}}(\tau') \right).
\end{equation}
Expanding the numerator, we obtain the following
\begin{equation*}
 \langle \hat{T}_\tau S(\beta, 0) \, \rho_c(\mathbf{q}, \tau) \rho_c(-\mathbf{q}, 0) \rangle_0 = \\
 \sum_{m=0}^{\infty} \frac{(V_c)^m}{m!}
\left\langle
\hat{T}_\tau \left( \int_0^{\beta} d\tau_1 \, H_{\text{CDW}}(\tau_1) \right)^m
\rho_c(\mathbf{q}, \tau) \rho_c(-\mathbf{q}, 0) \right\rangle_0,
\end{equation*}
which can be usefully interpreted in terms of connected and disconnected diagrams, as typically obtained via Wick's theorem~\cite{wicks, ma2024chargesusceptibility}. All disconnected-diagram-like terms are canceled by the denominator. To make this explicit, let us consider the first-order contribution, where the variable $\mathbf{q}$ is treated as an index to avoid cluttering the notation:
\begin{flalign*}
 \left\langle \hat{T}_\tau \, \rho_{-q}(\tau_1) \rho_q(\tau_1) \rho_q(\tau) \rho_{-q}(0) \right\rangle = 
 \quad \:\left\langle \hat{T}_\tau \, \rho_{-q}(\tau_1) \rho_q(\tau_1) \right\rangle 
  \left\langle \hat{T}_\tau \, \rho_q(\tau) \rho_{-q}(0) \right\rangle \\
 + \left\langle \hat{T}_\tau \, \rho_{-q}(\tau_1) \rho_q(\tau) \right\rangle 
  \left\langle \hat{T}_\tau \, \rho_q(\tau_1) \rho_{-q}(0) \right\rangle \\
 + \left\langle \hat{T}_\tau \, \rho_{-q}(\tau_1) \rho_{-q}(0) \right\rangle 
  \left\langle \hat{T}_\tau \, \rho_q(\tau_1) \rho_q(\tau) \right\rangle.
\end{flalign*}
Here, we note two important points. First, the term
\begin{equation*}
\left\langle \hat{T}_\tau \, \rho_{-q}(\tau_1) \rho_{-q}(0) \right\rangle 
\left\langle \hat{T}_\tau \, \rho_q(\tau_1) \rho_q(\tau) \right\rangle
\end{equation*}
can be interpreted in terms of a disconnected diagram, and contributes only when $\mathbf{q} = 0$. By contrast, if $\mathbf{q} \neq 0$ (as considered in the main text), this term vanishes.
For the two remaining terms, the product
\begin{equation*}
   \left\langle \hat{T}_\tau \,  \rho_{-q}(\tau_1) \rho_q(\tau) \right\rangle 
\left\langle \hat{T}_\tau \, \rho_q(\tau_1) \rho_{-q}(0) \right\rangle 
\end{equation*}
can be interpreted in terms of a connected-diagram term and yields a contribution of the form $\chi(\tau_1 - \tau)\chi(\tau_1)$. The other term is again disconnected. A similar structure persists at higher orders. For instance, at the next order, six non-zero terms arise (four disconnected and two connected diagram terms), resulting in the series
\begin{equation}
    \chi_c  = \chi_0 + V_c \chi_0^2 + (V_c)^2 \chi_{0}^3 + \cdots,
\end{equation}
where we now change the notation slightly and set $\chi_0\equiv \chi^{(0)}_c$. Summing all terms, we obtain the following result for the geometric series:
\begin{equation}
    \chi_c = \sum_{m=0}^\infty V_c^m \chi_0^{m+1}(i\nu_n) = \frac{\chi_0(i\nu_n)}{1 - V_c \chi_0(i\nu_n)}.
\end{equation}
This is the Dyson-like equation for the charge susceptibility of the present model, which is discussed in the main text.

\begin{figure}[t]
\centering
\centering \includegraphics[width=0.45\linewidth]{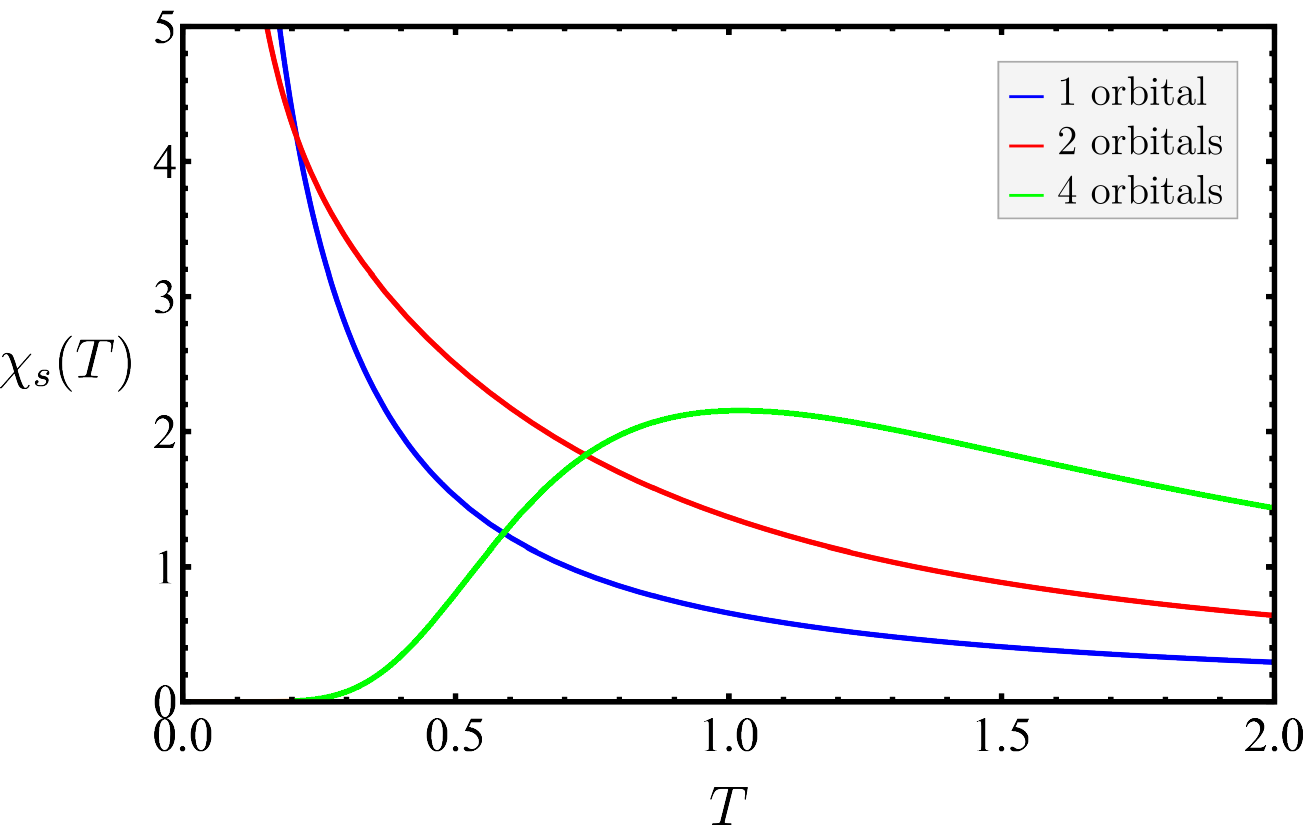}
\caption{Ferromagnetic susceptibility for $x=0.1$ and $u=1.5$ for different numbers of orbitals ($n$) included in the 2D OHK model. It is clear that beyond four orbitals ($n\geq 4$), the ferromagnetic susceptibility of the model becomes suppressed at low temperatures. Here, we assume units such that $\mu_B=1$.}
\label{orbs}
\end{figure}

\section{Ferromagnetic susceptibility in the 2D orbital HK model}
\label{Appendix_B}

The orbital extension of the 2D HK model (henceforth, denoted by OHK model) is particularly interesting. As shown in some recent works \cite{bradlyn2023, hackner2025solving, DynamicalPP2025, realspace2024,mai2024newapproachstrongcorr}, adding such degrees of freedom makes the 2D HK model more realistic since it can potentially capture the behavior displayed by the 2D Hubbard model defined with $N/n$ clusters ($N$ being the total number of sites). This offers a promising route for understanding some key features of the Hubbard model via an exactly solvable framework \cite{mai2024newapproachstrongcorr}, while also brings the corresponding effective HK model potentially closer to the realistic behavior of some strongly correlated materials.

The form of the extended Hamiltonian, now including $n$ orbitals, is given by \cite{bradlyn2023}
\begin{equation} \label{HOHK}
H_{\text{OHK}} = 
\sum_{\mathbf{k}, \alpha, \beta, \sigma} 
t_{\alpha \beta}(\mathbf{k}) \, c^\dagger_{\mathbf{k} \alpha \sigma} c_{\mathbf{k} \beta \sigma}
- \mu \sum_{\mathbf{k}, \alpha, \sigma} n_{\mathbf{k} \alpha \sigma}
+ \sum_{\mathbf{k}, \alpha, \beta} U_{\alpha \beta} \, n_{\mathbf{k} \alpha \uparrow} \, n_{\mathbf{k} \beta \downarrow}.
\end{equation}
As can be seen, the structure of this Hamiltonian is similar to the single-orbital case [Eq. \eqref{hamilto1}], but with the addition of the indices $\alpha$ and $\beta$, which label the different orbitals. The term $t_{\alpha \beta}(\mathbf{k})$ denotes the momentum-dependent hopping matrix between orbitals, while $U_{\alpha \beta}$ represents the interaction matrix.

Although this extension of the model may appear uneventful at first sight, it has a significant impact. A clear example is the Green's function in the OHK model, which can no longer be written in the same form as in the single-orbital case, thereby requiring the use of the Lehmann representation. This is because the occupation number operator $n_{\mathbf{k} \alpha, \sigma}$ no longer commutes with the Hamiltonian, and the structure of the creation and annihilation operators becomes non-trivial. Consequently, these operators must be evaluated in the eigenbasis of the interacting system, which can be understood more clearly by examining the following structure of the Green's function:
\begin{equation*}
\mathcal{G}_{\alpha\beta}^{\text{OHK}}(\mathbf{k}, \omega + i0^+)
= \frac{1}{Z} \sum_{m,n}
\frac{e^{-\beta E_{m}} + e^{-\beta E_{n}}}{\omega + i0^+ + E_{m} - E_{n}}
\, \langle m | \hat{c}_{\mathbf{k} \alpha \sigma} | n \rangle
\langle n | \hat{c}^{\dagger}_{\mathbf{k} \beta \sigma} | m \rangle,
\end{equation*}
where the matrix elements represent the creation and annihilation operators in the eigenbasis of the full Hamiltonian. Similarly, $E_{m}$ and $E_{n}$ are the corresponding many-body eigenenergies of the system. The partition function is given by ${Z} = \mathrm{Tr} \, e^{-\beta H_{\text{OHK}}}$, from which quantities such as the average occupation can be computed.

Observables such as the uniform spin susceptibility or pair susceptibility (both addressed in the present work) can be calculated using the same formalism, but now with a modified structure for the Green's function and partition function that incorporates the orbital degrees of freedom. As mentioned in the main text, an important result that we want to emphasize here is that by increasing the number of orbitals in the 2D OHK model, this leads to a suppression of the ferromagnetic susceptibility near half-filling. This behavior is clearly observed in Fig.~\ref{orbs}. Since the 2D Hubbard model is expected to be described by the OHK model in the large-$n$ limit, this result is consistent with the expected behavior of the Hubbard model in that regime.


\begin{thebibliography}{49}%
\makeatletter
\providecommand \@ifxundefined [1]{%
 \@ifx{#1\undefined}
}%
\providecommand \@ifnum [1]{%
 \ifnum #1\expandafter \@firstoftwo
 \else \expandafter \@secondoftwo
 \fi
}%
\providecommand \@ifx [1]{%
 \ifx #1\expandafter \@firstoftwo
 \else \expandafter \@secondoftwo
 \fi
}%
\providecommand \natexlab [1]{#1}%
\providecommand \enquote  [1]{``#1''}%
\providecommand \bibnamefont  [1]{#1}%
\providecommand \bibfnamefont [1]{#1}%
\providecommand \citenamefont [1]{#1}%
\providecommand \href@noop [0]{\@secondoftwo}%
\providecommand \href [0]{\begingroup \@sanitize@url \@href}%
\providecommand \@href[1]{\@@startlink{#1}\@@href}%
\providecommand \@@href[1]{\endgroup#1\@@endlink}%
\providecommand \@sanitize@url [0]{\catcode `\\12\catcode `\$12\catcode
  `\&12\catcode `\#12\catcode `\^12\catcode `\_12\catcode `\%12\relax}%
\providecommand \@@startlink[1]{}%
\providecommand \@@endlink[0]{}%
\providecommand \url  [0]{\begingroup\@sanitize@url \@url }%
\providecommand \@url [1]{\endgroup\@href {#1}{\urlprefix }}%
\providecommand \urlprefix  [0]{URL }%
\providecommand \Eprint [0]{\href }%
\providecommand \doibase [0]{https://doi.org/}%
\providecommand \selectlanguage [0]{\@gobble}%
\providecommand \bibinfo  [0]{\@secondoftwo}%
\providecommand \bibfield  [0]{\@secondoftwo}%
\providecommand \translation [1]{[#1]}%
\providecommand \BibitemOpen [0]{}%
\providecommand \bibitemStop [0]{}%
\providecommand \bibitemNoStop [0]{.\EOS\space}%
\providecommand \EOS [0]{\spacefactor3000\relax}%
\providecommand \BibitemShut  [1]{\csname bibitem#1\endcsname}%
\let\auto@bib@innerbib\@empty
\bibitem [{\citenamefont {Keimer}\ \emph {et~al.}(2015)\citenamefont {Keimer},
  \citenamefont {Kivelson}, \citenamefont {Norman}, \citenamefont {Uchida},\
  and\ \citenamefont {Zaanen}}]{keimer_review}%
  \BibitemOpen
  \bibfield  {author} {\bibinfo {author} {\bibfnamefont {B.}~\bibnamefont
  {Keimer}}, \bibinfo {author} {\bibfnamefont {S.~A.}\ \bibnamefont
  {Kivelson}}, \bibinfo {author} {\bibfnamefont {M.~R.}\ \bibnamefont
  {Norman}}, \bibinfo {author} {\bibfnamefont {S.}~\bibnamefont {Uchida}},\
  and\ \bibinfo {author} {\bibfnamefont {J.}~\bibnamefont {Zaanen}},\ }\href
  {https://www.osti.gov/biblio/1357580} {\bibfield  {journal} {\bibinfo
  {journal} {Nature (London)}\ }\textbf {\bibinfo {volume} {518}} (\bibinfo
  {year} {2015})}\BibitemShut {NoStop}%
\bibitem [{\citenamefont {Lee}\ \emph {et~al.}(2006)\citenamefont {Lee},
  \citenamefont {Nagaosa},\ and\ \citenamefont {Wen}}]{PALee_review}%
  \BibitemOpen
  \bibfield  {author} {\bibinfo {author} {\bibfnamefont {P.~A.}\ \bibnamefont
  {Lee}}, \bibinfo {author} {\bibfnamefont {N.}~\bibnamefont {Nagaosa}},\ and\
  \bibinfo {author} {\bibfnamefont {X.-G.}\ \bibnamefont {Wen}},\ }\href
  {https://doi.org/10.1103/RevModPhys.78.17} {\bibfield  {journal} {\bibinfo
  {journal} {Rev. Mod. Phys.}\ }\textbf {\bibinfo {volume} {78}},\ \bibinfo
  {pages} {17} (\bibinfo {year} {2006})}\BibitemShut {NoStop}%
\bibitem [{\citenamefont {Anderson}(1987)}]{anderson1987resonating}%
  \BibitemOpen
  \bibfield  {author} {\bibinfo {author} {\bibfnamefont {P.~W.}\ \bibnamefont
  {Anderson}},\ }\href@noop {} {\bibfield  {journal} {\bibinfo  {journal}
  {Science}\ }\textbf {\bibinfo {volume} {235}},\ \bibinfo {pages} {1196}
  (\bibinfo {year} {1987})}\BibitemShut {NoStop}%
\bibitem [{\citenamefont {LeBlanc}\ \emph {et~al.}(2015)\citenamefont
  {LeBlanc}, \citenamefont {Antipov}, \citenamefont {Becca}, \citenamefont
  {Bulik}, \citenamefont {Chan}, \citenamefont {Chung}, \citenamefont {Deng},
  \citenamefont {Ferrero}, \citenamefont {Henderson}, \citenamefont
  {Jim\'enez-Hoyos}, \citenamefont {Kozik}, \citenamefont {Liu}, \citenamefont
  {Millis}, \citenamefont {Prokof'ev}, \citenamefont {Qin}, \citenamefont
  {Scuseria}, \citenamefont {Shi}, \citenamefont {Svistunov}, \citenamefont
  {Tocchio}, \citenamefont {Tupitsyn}, \citenamefont {White}, \citenamefont
  {Zhang}, \citenamefont {Zheng}, \citenamefont {Zhu},\ and\ \citenamefont
  {Gull}}]{leblanc2015solutions}%
  \BibitemOpen
  \bibfield  {author} {\bibinfo {author} {\bibfnamefont {J.~P.~F.}\
  \bibnamefont {LeBlanc}}, \bibinfo {author} {\bibfnamefont {A.~E.}\
  \bibnamefont {Antipov}}, \bibinfo {author} {\bibfnamefont {F.}~\bibnamefont
  {Becca}}, \bibinfo {author} {\bibfnamefont {I.~W.}\ \bibnamefont {Bulik}},
  \bibinfo {author} {\bibfnamefont {G.~K.-L.}\ \bibnamefont {Chan}}, \bibinfo
  {author} {\bibfnamefont {C.-M.}\ \bibnamefont {Chung}}, \bibinfo {author}
  {\bibfnamefont {Y.}~\bibnamefont {Deng}}, \bibinfo {author} {\bibfnamefont
  {M.}~\bibnamefont {Ferrero}}, \bibinfo {author} {\bibfnamefont {T.~M.}\
  \bibnamefont {Henderson}}, \bibinfo {author} {\bibfnamefont {C.~A.}\
  \bibnamefont {Jim\'enez-Hoyos}}, \bibinfo {author} {\bibfnamefont
  {E.}~\bibnamefont {Kozik}}, \bibinfo {author} {\bibfnamefont {X.-W.}\
  \bibnamefont {Liu}}, \bibinfo {author} {\bibfnamefont {A.~J.}\ \bibnamefont
  {Millis}}, \bibinfo {author} {\bibfnamefont {N.~V.}\ \bibnamefont
  {Prokof'ev}}, \bibinfo {author} {\bibfnamefont {M.}~\bibnamefont {Qin}},
  \bibinfo {author} {\bibfnamefont {G.~E.}\ \bibnamefont {Scuseria}}, \bibinfo
  {author} {\bibfnamefont {H.}~\bibnamefont {Shi}}, \bibinfo {author}
  {\bibfnamefont {B.~V.}\ \bibnamefont {Svistunov}}, \bibinfo {author}
  {\bibfnamefont {L.~F.}\ \bibnamefont {Tocchio}}, \bibinfo {author}
  {\bibfnamefont {I.~S.}\ \bibnamefont {Tupitsyn}}, \bibinfo {author}
  {\bibfnamefont {S.~R.}\ \bibnamefont {White}}, \bibinfo {author}
  {\bibfnamefont {S.}~\bibnamefont {Zhang}}, \bibinfo {author} {\bibfnamefont
  {B.-X.}\ \bibnamefont {Zheng}}, \bibinfo {author} {\bibfnamefont
  {Z.}~\bibnamefont {Zhu}},\ and\ \bibinfo {author} {\bibfnamefont
  {E.}~\bibnamefont {Gull}} (\bibinfo {collaboration} {Simons Collaboration on
  the Many-Electron Problem}),\ }\href
  {https://doi.org/10.1103/PhysRevX.5.041041} {\bibfield  {journal} {\bibinfo
  {journal} {Phys. Rev. X}\ }\textbf {\bibinfo {volume} {5}},\ \bibinfo {pages}
  {041041} (\bibinfo {year} {2015})}\BibitemShut {NoStop}%
\bibitem [{\citenamefont {Arovas}\ \emph {et~al.}(2022)\citenamefont {Arovas},
  \citenamefont {Berg}, \citenamefont {Kivelson},\ and\ \citenamefont
  {Raghu}}]{arovas2022hubbard}%
  \BibitemOpen
  \bibfield  {author} {\bibinfo {author} {\bibfnamefont {D.~P.}\ \bibnamefont
  {Arovas}}, \bibinfo {author} {\bibfnamefont {E.}~\bibnamefont {Berg}},
  \bibinfo {author} {\bibfnamefont {S.~A.}\ \bibnamefont {Kivelson}},\ and\
  \bibinfo {author} {\bibfnamefont {S.}~\bibnamefont {Raghu}},\ }\href@noop {}
  {\bibfield  {journal} {\bibinfo  {journal} {Annual Review of Condensed Matter
  Physics}\ }\textbf {\bibinfo {volume} {13}},\ \bibinfo {pages} {239}
  (\bibinfo {year} {2022})}\BibitemShut {NoStop}%
\bibitem [{\citenamefont {Hatsugai}\ and\ \citenamefont
  {Kohmoto}(1992)}]{hatsugai1992exactly}%
  \BibitemOpen
  \bibfield  {author} {\bibinfo {author} {\bibfnamefont {Y.}~\bibnamefont
  {Hatsugai}}\ and\ \bibinfo {author} {\bibfnamefont {M.}~\bibnamefont
  {Kohmoto}},\ }\href@noop {} {\bibfield  {journal} {\bibinfo  {journal}
  {Journal of the Physical Society of Japan}\ }\textbf {\bibinfo {volume}
  {61}},\ \bibinfo {pages} {2056} (\bibinfo {year} {1992})}\BibitemShut
  {NoStop}%
\bibitem [{\citenamefont {{Baskaran}}(1991)}]{G_Baskaran}%
  \BibitemOpen
  \bibfield  {author} {\bibinfo {author} {\bibfnamefont {G.}~\bibnamefont
  {{Baskaran}}},\ }\href {https://doi.org/10.1142/S0217984991000782} {\bibfield
   {journal} {\bibinfo  {journal} {Modern Physics Letters B}\ }\textbf
  {\bibinfo {volume} {5}},\ \bibinfo {pages} {643} (\bibinfo {year}
  {1991})}\BibitemShut {NoStop}%
\bibitem [{\citenamefont {Continentino}\ and\ \citenamefont
  {Coutinho-Filho}(1994)}]{continentino1994scaling}%
  \BibitemOpen
  \bibfield  {author} {\bibinfo {author} {\bibfnamefont {M.~A.}\ \bibnamefont
  {Continentino}}\ and\ \bibinfo {author} {\bibfnamefont {M.~D.}\ \bibnamefont
  {Coutinho-Filho}},\ }\href@noop {} {\bibfield  {journal} {\bibinfo  {journal}
  {Solid State Communications}\ }\textbf {\bibinfo {volume} {90}},\ \bibinfo
  {pages} {619} (\bibinfo {year} {1994})}\BibitemShut {NoStop}%
\bibitem [{\citenamefont {Vitoriano}\ \emph {et~al.}(2000)\citenamefont
  {Vitoriano}, \citenamefont {Bejan}, \citenamefont {Mac{\^e}do},\ and\
  \citenamefont {Coutinho-Filho}}]{vitoriano2000metal}%
  \BibitemOpen
  \bibfield  {author} {\bibinfo {author} {\bibfnamefont {C.}~\bibnamefont
  {Vitoriano}}, \bibinfo {author} {\bibfnamefont {L.}~\bibnamefont {Bejan}},
  \bibinfo {author} {\bibfnamefont {A.}~\bibnamefont {Mac{\^e}do}},\ and\
  \bibinfo {author} {\bibfnamefont {M.}~\bibnamefont {Coutinho-Filho}},\
  }\href@noop {} {\bibfield  {journal} {\bibinfo  {journal} {Physical Review
  B}\ }\textbf {\bibinfo {volume} {61}},\ \bibinfo {pages} {7941} (\bibinfo
  {year} {2000})}\BibitemShut {NoStop}%
\bibitem [{\citenamefont {Yeo}\ and\ \citenamefont
  {Phillips}(2019)}]{yeo2019local}%
  \BibitemOpen
  \bibfield  {author} {\bibinfo {author} {\bibfnamefont {L.}~\bibnamefont
  {Yeo}}\ and\ \bibinfo {author} {\bibfnamefont {P.~W.}\ \bibnamefont
  {Phillips}},\ }\href@noop {} {\bibfield  {journal} {\bibinfo  {journal}
  {Physical Review D}\ }\textbf {\bibinfo {volume} {99}},\ \bibinfo {pages}
  {094030} (\bibinfo {year} {2019})}\BibitemShut {NoStop}%
\bibitem [{\citenamefont {Phillips}\ \emph {et~al.}(2020)\citenamefont
  {Phillips}, \citenamefont {Yeo},\ and\ \citenamefont {Huang}}]{Phillips2020}%
  \BibitemOpen
  \bibfield  {author} {\bibinfo {author} {\bibfnamefont {P.~W.}\ \bibnamefont
  {Phillips}}, \bibinfo {author} {\bibfnamefont {L.}~\bibnamefont {Yeo}},\ and\
  \bibinfo {author} {\bibfnamefont {E.~W.}\ \bibnamefont {Huang}},\ }\href
  {https://doi.org/10.1038/s41567-020-0988-4} {\bibfield  {journal} {\bibinfo
  {journal} {Nature Physics}\ }\textbf {\bibinfo {volume} {16}},\ \bibinfo
  {pages} {1175} (\bibinfo {year} {2020})}\BibitemShut {NoStop}%
\bibitem [{\citenamefont {Li}\ \emph {et~al.}(2022)\citenamefont {Li},
  \citenamefont {Mishra}, \citenamefont {Zhou},\ and\ \citenamefont
  {Zhang}}]{li2022two}%
  \BibitemOpen
  \bibfield  {author} {\bibinfo {author} {\bibfnamefont {Y.}~\bibnamefont
  {Li}}, \bibinfo {author} {\bibfnamefont {V.}~\bibnamefont {Mishra}}, \bibinfo
  {author} {\bibfnamefont {Y.}~\bibnamefont {Zhou}},\ and\ \bibinfo {author}
  {\bibfnamefont {F.-C.}\ \bibnamefont {Zhang}},\ }\href@noop {} {\bibfield
  {journal} {\bibinfo  {journal} {New Journal of Physics}\ }\textbf {\bibinfo
  {volume} {24}},\ \bibinfo {pages} {103019} (\bibinfo {year}
  {2022})}\BibitemShut {NoStop}%
\bibitem [{\citenamefont {Fradkin}\ \emph {et~al.}(2015)\citenamefont
  {Fradkin}, \citenamefont {Kivelson},\ and\ \citenamefont
  {Tranquada}}]{fradkin2015colloquium}%
  \BibitemOpen
  \bibfield  {author} {\bibinfo {author} {\bibfnamefont {E.}~\bibnamefont
  {Fradkin}}, \bibinfo {author} {\bibfnamefont {S.~A.}\ \bibnamefont
  {Kivelson}},\ and\ \bibinfo {author} {\bibfnamefont {J.~M.}\ \bibnamefont
  {Tranquada}},\ }\href {https://doi.org/10.1103/RevModPhys.87.457} {\bibfield
  {journal} {\bibinfo  {journal} {Reviews of Modern Physics}\ }\textbf
  {\bibinfo {volume} {87}},\ \bibinfo {pages} {457} (\bibinfo {year}
  {2015})}\BibitemShut {NoStop}%
\bibitem [{\citenamefont {Timusk}\ and\ \citenamefont
  {Statt}(1999)}]{timusk1999pseudogap}%
  \BibitemOpen
  \bibfield  {author} {\bibinfo {author} {\bibfnamefont {T.}~\bibnamefont
  {Timusk}}\ and\ \bibinfo {author} {\bibfnamefont {B.}~\bibnamefont {Statt}},\
  }\href {https://doi.org/10.1088/0034-4885/62/1/002} {\bibfield  {journal}
  {\bibinfo  {journal} {Reports on Progress in Physics}\ }\textbf {\bibinfo
  {volume} {62}},\ \bibinfo {pages} {61} (\bibinfo {year} {1999})}\BibitemShut
  {NoStop}%
\bibitem [{\citenamefont {Yang}(2021)}]{fermiarcs}%
  \BibitemOpen
  \bibfield  {author} {\bibinfo {author} {\bibfnamefont {K.}~\bibnamefont
  {Yang}},\ }\href {https://doi.org/10.1103/PhysRevB.103.024529} {\bibfield
  {journal} {\bibinfo  {journal} {Phys. Rev. B}\ }\textbf {\bibinfo {volume}
  {103}},\ \bibinfo {pages} {024529} (\bibinfo {year} {2021})}\BibitemShut
  {NoStop}%
\bibitem [{\citenamefont {Zhu}\ and\ \citenamefont
  {Han}(2021)}]{zhu2021effects}%
  \BibitemOpen
  \bibfield  {author} {\bibinfo {author} {\bibfnamefont {H.-S.}\ \bibnamefont
  {Zhu}}\ and\ \bibinfo {author} {\bibfnamefont {Q.}~\bibnamefont {Han}},\
  }\href@noop {} {\bibfield  {journal} {\bibinfo  {journal} {Chinese Physics
  B}\ }\textbf {\bibinfo {volume} {30}},\ \bibinfo {pages} {107401} (\bibinfo
  {year} {2021})}\BibitemShut {NoStop}%
\bibitem [{\citenamefont {Froldi}\ \emph {et~al.}(2024)\citenamefont {Froldi},
  \citenamefont {Corsino},\ and\ \citenamefont {Freire}}]{froldi2024strong}%
  \BibitemOpen
  \bibfield  {author} {\bibinfo {author} {\bibfnamefont {I.~d.~M.}\
  \bibnamefont {Froldi}}, \bibinfo {author} {\bibfnamefont {C.~E. S.~P.}\
  \bibnamefont {Corsino}},\ and\ \bibinfo {author} {\bibfnamefont
  {H.}~\bibnamefont {Freire}},\ }\href
  {https://doi.org/10.1103/PhysRevB.110.245136} {\bibfield  {journal} {\bibinfo
   {journal} {Phys. Rev. B}\ }\textbf {\bibinfo {volume} {110}},\ \bibinfo
  {pages} {245136} (\bibinfo {year} {2024})}\BibitemShut {NoStop}%
\bibitem [{\citenamefont {Zhao}\ \emph {et~al.}(2025)\citenamefont {Zhao},
  \citenamefont {Yang},\ and\ \citenamefont {Zhong}}]{Zhao_review_2025}%
  \BibitemOpen
  \bibfield  {author} {\bibinfo {author} {\bibfnamefont {M.}~\bibnamefont
  {Zhao}}, \bibinfo {author} {\bibfnamefont {W.-W.}\ \bibnamefont {Yang}},\
  and\ \bibinfo {author} {\bibfnamefont {Y.}~\bibnamefont {Zhong}},\ }\href
  {https://doi.org/10.1088/1361-648X/adc64c} {\bibfield  {journal} {\bibinfo
  {journal} {Journal of Physics: Condensed Matter}\ }\textbf {\bibinfo {volume}
  {37}},\ \bibinfo {pages} {183005} (\bibinfo {year} {2025})}\BibitemShut
  {NoStop}%
\bibitem [{\citenamefont {Cohen-Stead}\ \emph {et~al.}(2019)\citenamefont
  {Cohen-Stead}, \citenamefont {Costa}, \citenamefont {Khatami},\ and\
  \citenamefont {Scalettar}}]{Scalettar_strain}%
  \BibitemOpen
  \bibfield  {author} {\bibinfo {author} {\bibfnamefont {B.}~\bibnamefont
  {Cohen-Stead}}, \bibinfo {author} {\bibfnamefont {N.~C.}\ \bibnamefont
  {Costa}}, \bibinfo {author} {\bibfnamefont {E.}~\bibnamefont {Khatami}},\
  and\ \bibinfo {author} {\bibfnamefont {R.~T.}\ \bibnamefont {Scalettar}},\
  }\href {https://doi.org/10.1103/PhysRevB.100.045125} {\bibfield  {journal}
  {\bibinfo  {journal} {Phys. Rev. B}\ }\textbf {\bibinfo {volume} {100}},\
  \bibinfo {pages} {045125} (\bibinfo {year} {2019})}\BibitemShut {NoStop}%
\bibitem [{\citenamefont {Cooper}(1956)}]{cooperoriginal}%
  \BibitemOpen
  \bibfield  {author} {\bibinfo {author} {\bibfnamefont {L.~N.}\ \bibnamefont
  {Cooper}},\ }\href {https://doi.org/10.1103/PhysRev.104.1189} {\bibfield
  {journal} {\bibinfo  {journal} {Phys. Rev.}\ }\textbf {\bibinfo {volume}
  {104}},\ \bibinfo {pages} {1189} (\bibinfo {year} {1956})}\BibitemShut
  {NoStop}%
\bibitem [{\citenamefont {Bardeen}\ \emph {et~al.}(1957)\citenamefont
  {Bardeen}, \citenamefont {Cooper},\ and\ \citenamefont
  {Schrieffer}}]{bcs1957}%
  \BibitemOpen
  \bibfield  {author} {\bibinfo {author} {\bibfnamefont {J.}~\bibnamefont
  {Bardeen}}, \bibinfo {author} {\bibfnamefont {L.~N.}\ \bibnamefont
  {Cooper}},\ and\ \bibinfo {author} {\bibfnamefont {J.~R.}\ \bibnamefont
  {Schrieffer}},\ }\href {https://doi.org/10.1103/PhysRev.108.1175} {\bibfield
  {journal} {\bibinfo  {journal} {Phys. Rev.}\ }\textbf {\bibinfo {volume}
  {108}},\ \bibinfo {pages} {1175} (\bibinfo {year} {1957})}\BibitemShut
  {NoStop}%
\bibitem [{\citenamefont {Scalapino}(1995)}]{scalapino1995case}%
  \BibitemOpen
  \bibfield  {author} {\bibinfo {author} {\bibfnamefont {D.~J.}\ \bibnamefont
  {Scalapino}},\ }\href@noop {} {\bibfield  {journal} {\bibinfo  {journal}
  {Physics Reports}\ }\textbf {\bibinfo {volume} {250}},\ \bibinfo {pages}
  {329} (\bibinfo {year} {1995})}\BibitemShut {NoStop}%
\bibitem [{\citenamefont {Sigrist}\ and\ \citenamefont {Ueda}(1991)}]{pwave1}%
  \BibitemOpen
  \bibfield  {author} {\bibinfo {author} {\bibfnamefont {M.}~\bibnamefont
  {Sigrist}}\ and\ \bibinfo {author} {\bibfnamefont {K.}~\bibnamefont {Ueda}},\
  }\href {https://doi.org/10.1103/RevModPhys.63.239} {\bibfield  {journal}
  {\bibinfo  {journal} {Rev. Mod. Phys.}\ }\textbf {\bibinfo {volume} {63}},\
  \bibinfo {pages} {239} (\bibinfo {year} {1991})}\BibitemShut {NoStop}%
\bibitem [{\citenamefont {Guerci}\ \emph {et~al.}(2025)\citenamefont {Guerci},
  \citenamefont {Sangiovanni}, \citenamefont {Millis},\ and\ \citenamefont
  {Fabrizio}}]{guerciHK}%
  \BibitemOpen
  \bibfield  {author} {\bibinfo {author} {\bibfnamefont {D.}~\bibnamefont
  {Guerci}}, \bibinfo {author} {\bibfnamefont {G.}~\bibnamefont {Sangiovanni}},
  \bibinfo {author} {\bibfnamefont {A.~J.}\ \bibnamefont {Millis}},\ and\
  \bibinfo {author} {\bibfnamefont {M.}~\bibnamefont {Fabrizio}},\ }\href
  {https://doi.org/10.1103/PhysRevB.111.075124} {\bibfield  {journal} {\bibinfo
   {journal} {Phys. Rev. B}\ }\textbf {\bibinfo {volume} {111}},\ \bibinfo
  {pages} {075124} (\bibinfo {year} {2025})}\BibitemShut {NoStop}%
\bibitem [{\citenamefont {Freire}\ \emph {et~al.}(2008)\citenamefont {Freire},
  \citenamefont {Correa},\ and\ \citenamefont {Ferraz}}]{Freire_Ferraz_2008}%
  \BibitemOpen
  \bibfield  {author} {\bibinfo {author} {\bibfnamefont {H.}~\bibnamefont
  {Freire}}, \bibinfo {author} {\bibfnamefont {E.}~\bibnamefont {Correa}},\
  and\ \bibinfo {author} {\bibfnamefont {A.}~\bibnamefont {Ferraz}},\ }\href
  {https://doi.org/10.1103/PhysRevB.78.125114} {\bibfield  {journal} {\bibinfo
  {journal} {Phys. Rev. B}\ }\textbf {\bibinfo {volume} {78}},\ \bibinfo
  {pages} {125114} (\bibinfo {year} {2008})}\BibitemShut {NoStop}%
\bibitem [{\citenamefont {Correa}\ \emph {et~al.}(2008)\citenamefont {Correa},
  \citenamefont {Freire},\ and\ \citenamefont
  {Ferraz}}]{Correa_Freire_Ferraz_2008}%
  \BibitemOpen
  \bibfield  {author} {\bibinfo {author} {\bibfnamefont {E.}~\bibnamefont
  {Correa}}, \bibinfo {author} {\bibfnamefont {H.}~\bibnamefont {Freire}},\
  and\ \bibinfo {author} {\bibfnamefont {A.}~\bibnamefont {Ferraz}},\ }\href
  {https://doi.org/10.1103/PhysRevB.78.195108} {\bibfield  {journal} {\bibinfo
  {journal} {Phys. Rev. B}\ }\textbf {\bibinfo {volume} {78}},\ \bibinfo
  {pages} {195108} (\bibinfo {year} {2008})}\BibitemShut {NoStop}%
\bibitem [{\citenamefont {Freire}\ \emph {et~al.}(2005)\citenamefont {Freire},
  \citenamefont {Corr\^ea},\ and\ \citenamefont {Ferraz}}]{Freire_Ferraz_2005}%
  \BibitemOpen
  \bibfield  {author} {\bibinfo {author} {\bibfnamefont {H.}~\bibnamefont
  {Freire}}, \bibinfo {author} {\bibfnamefont {E.}~\bibnamefont {Corr\^ea}},\
  and\ \bibinfo {author} {\bibfnamefont {A.}~\bibnamefont {Ferraz}},\ }\href
  {https://doi.org/10.1103/PhysRevB.71.165113} {\bibfield  {journal} {\bibinfo
  {journal} {Phys. Rev. B}\ }\textbf {\bibinfo {volume} {71}},\ \bibinfo
  {pages} {165113} (\bibinfo {year} {2005})}\BibitemShut {NoStop}%
\bibitem [{\citenamefont {Caetano}\ and\ \citenamefont
  {Freire}(2019)}]{Freire_Caetano_2019}%
  \BibitemOpen
  \bibfield  {author} {\bibinfo {author} {\bibfnamefont {R.~R.}\ \bibnamefont
  {Caetano}}\ and\ \bibinfo {author} {\bibfnamefont {H.}~\bibnamefont
  {Freire}},\ }\href {https://doi.org/10.1016/j.aop.2019.03.024} {\bibfield
  {journal} {\bibinfo  {journal} {Annals of Physics}\ }\textbf {\bibinfo
  {volume} {405}},\ \bibinfo {pages} {308–324} (\bibinfo {year}
  {2019})}\BibitemShut {NoStop}%
\bibitem [{\citenamefont {Agterberg}\ \emph {et~al.}(2020)\citenamefont
  {Agterberg}, \citenamefont {Davis}, \citenamefont {Edkins}, \citenamefont
  {Fradkin}, \citenamefont {Van~Harlingen}, \citenamefont {Kivelson},
  \citenamefont {Lee}, \citenamefont {Radzihovsky}, \citenamefont {Tranquada},\
  and\ \citenamefont {Wang}}]{Agterberg_2020}%
  \BibitemOpen
  \bibfield  {author} {\bibinfo {author} {\bibfnamefont {D.~F.}\ \bibnamefont
  {Agterberg}}, \bibinfo {author} {\bibfnamefont {J.~S.}\ \bibnamefont
  {Davis}}, \bibinfo {author} {\bibfnamefont {S.~D.}\ \bibnamefont {Edkins}},
  \bibinfo {author} {\bibfnamefont {E.}~\bibnamefont {Fradkin}}, \bibinfo
  {author} {\bibfnamefont {D.~J.}\ \bibnamefont {Van~Harlingen}}, \bibinfo
  {author} {\bibfnamefont {S.~A.}\ \bibnamefont {Kivelson}}, \bibinfo {author}
  {\bibfnamefont {P.~A.}\ \bibnamefont {Lee}}, \bibinfo {author} {\bibfnamefont
  {L.}~\bibnamefont {Radzihovsky}}, \bibinfo {author} {\bibfnamefont {J.~M.}\
  \bibnamefont {Tranquada}},\ and\ \bibinfo {author} {\bibfnamefont
  {Y.}~\bibnamefont {Wang}},\ }\href
  {https://doi.org/10.1146/annurev-conmatphys-031119-050711} {\bibfield
  {journal} {\bibinfo  {journal} {Annual Review of Condensed Matter Physics}\
  }\textbf {\bibinfo {volume} {11}},\ \bibinfo {pages} {231} (\bibinfo {year}
  {2020})}\BibitemShut {NoStop}%
\bibitem [{\citenamefont {de~Carvalho}\ and\ \citenamefont
  {Freire}(2014)}]{de_Carvalho_2014}%
  \BibitemOpen
  \bibfield  {author} {\bibinfo {author} {\bibfnamefont {V.~S.}\ \bibnamefont
  {de~Carvalho}}\ and\ \bibinfo {author} {\bibfnamefont {H.}~\bibnamefont
  {Freire}},\ }\href {https://doi.org/10.1016/j.aop.2014.05.009} {\bibfield
  {journal} {\bibinfo  {journal} {Annals of Physics}\ }\textbf {\bibinfo
  {volume} {348}},\ \bibinfo {pages} {32–49} (\bibinfo {year}
  {2014})}\BibitemShut {NoStop}%
\bibitem [{\citenamefont {Freire}\ \emph {et~al.}(2015)\citenamefont {Freire},
  \citenamefont {de~Carvalho},\ and\ \citenamefont {P\'epin}}]{Freire_2015}%
  \BibitemOpen
  \bibfield  {author} {\bibinfo {author} {\bibfnamefont {H.}~\bibnamefont
  {Freire}}, \bibinfo {author} {\bibfnamefont {V.~S.}\ \bibnamefont
  {de~Carvalho}},\ and\ \bibinfo {author} {\bibfnamefont {C.}~\bibnamefont
  {P\'epin}},\ }\href {https://doi.org/10.1103/PhysRevB.92.045132} {\bibfield
  {journal} {\bibinfo  {journal} {Phys. Rev. B}\ }\textbf {\bibinfo {volume}
  {92}},\ \bibinfo {pages} {045132} (\bibinfo {year} {2015})}\BibitemShut
  {NoStop}%
\bibitem [{\citenamefont {de~Carvalho}\ \emph {et~al.}(2015)\citenamefont
  {de~Carvalho}, \citenamefont {Kloss}, \citenamefont {Montiel}, \citenamefont
  {Freire},\ and\ \citenamefont {P\'epin}}]{Freire_loop_current_2015}%
  \BibitemOpen
  \bibfield  {author} {\bibinfo {author} {\bibfnamefont {V.~S.}\ \bibnamefont
  {de~Carvalho}}, \bibinfo {author} {\bibfnamefont {T.}~\bibnamefont {Kloss}},
  \bibinfo {author} {\bibfnamefont {X.}~\bibnamefont {Montiel}}, \bibinfo
  {author} {\bibfnamefont {H.}~\bibnamefont {Freire}},\ and\ \bibinfo {author}
  {\bibfnamefont {C.}~\bibnamefont {P\'epin}},\ }\href
  {https://doi.org/10.1103/PhysRevB.92.075123} {\bibfield  {journal} {\bibinfo
  {journal} {Phys. Rev. B}\ }\textbf {\bibinfo {volume} {92}},\ \bibinfo
  {pages} {075123} (\bibinfo {year} {2015})}\BibitemShut {NoStop}%
\bibitem [{\citenamefont {de~Carvalho}\ \emph {et~al.}(2016)\citenamefont
  {de~Carvalho}, \citenamefont {P\'epin},\ and\ \citenamefont
  {Freire}}]{Freire_loop_current_2016}%
  \BibitemOpen
  \bibfield  {author} {\bibinfo {author} {\bibfnamefont {V.~S.}\ \bibnamefont
  {de~Carvalho}}, \bibinfo {author} {\bibfnamefont {C.}~\bibnamefont
  {P\'epin}},\ and\ \bibinfo {author} {\bibfnamefont {H.}~\bibnamefont
  {Freire}},\ }\href {https://doi.org/10.1103/PhysRevB.93.115144} {\bibfield
  {journal} {\bibinfo  {journal} {Phys. Rev. B}\ }\textbf {\bibinfo {volume}
  {93}},\ \bibinfo {pages} {115144} (\bibinfo {year} {2016})}\BibitemShut
  {NoStop}%
\bibitem [{\citenamefont {Kloss}\ \emph {et~al.}(2016)\citenamefont {Kloss},
  \citenamefont {Montiel}, \citenamefont {de~Carvalho}, \citenamefont
  {Freire},\ and\ \citenamefont {Pépin}}]{Kloss_2016}%
  \BibitemOpen
  \bibfield  {author} {\bibinfo {author} {\bibfnamefont {T.}~\bibnamefont
  {Kloss}}, \bibinfo {author} {\bibfnamefont {X.}~\bibnamefont {Montiel}},
  \bibinfo {author} {\bibfnamefont {V.~S.}\ \bibnamefont {de~Carvalho}},
  \bibinfo {author} {\bibfnamefont {H.}~\bibnamefont {Freire}},\ and\ \bibinfo
  {author} {\bibfnamefont {C.}~\bibnamefont {Pépin}},\ }\href
  {https://doi.org/10.1088/0034-4885/79/8/084507} {\bibfield  {journal}
  {\bibinfo  {journal} {Reports on Progress in Physics}\ }\textbf {\bibinfo
  {volume} {79}},\ \bibinfo {pages} {084507} (\bibinfo {year}
  {2016})}\BibitemShut {NoStop}%
\bibitem [{\citenamefont {Hamidian}\ \emph {et~al.}(2016)\citenamefont
  {Hamidian}, \citenamefont {Edkins}, \citenamefont {Joo}, \citenamefont
  {Kostin}, \citenamefont {Eisaki}, \citenamefont {Uchida}, \citenamefont
  {Lawler}, \citenamefont {Kim}, \citenamefont {Mackenzie}, \citenamefont
  {Fujita}, \citenamefont {Lee},\ and\ \citenamefont {Davis}}]{Hamidian_2016}%
  \BibitemOpen
  \bibfield  {author} {\bibinfo {author} {\bibfnamefont {M.~H.}\ \bibnamefont
  {Hamidian}}, \bibinfo {author} {\bibfnamefont {S.~D.}\ \bibnamefont
  {Edkins}}, \bibinfo {author} {\bibfnamefont {S.~H.}\ \bibnamefont {Joo}},
  \bibinfo {author} {\bibfnamefont {A.}~\bibnamefont {Kostin}}, \bibinfo
  {author} {\bibfnamefont {H.}~\bibnamefont {Eisaki}}, \bibinfo {author}
  {\bibfnamefont {S.}~\bibnamefont {Uchida}}, \bibinfo {author} {\bibfnamefont
  {M.~J.}\ \bibnamefont {Lawler}}, \bibinfo {author} {\bibfnamefont {E.-A.}\
  \bibnamefont {Kim}}, \bibinfo {author} {\bibfnamefont {A.~P.}\ \bibnamefont
  {Mackenzie}}, \bibinfo {author} {\bibfnamefont {K.}~\bibnamefont {Fujita}},
  \bibinfo {author} {\bibfnamefont {J.}~\bibnamefont {Lee}},\ and\ \bibinfo
  {author} {\bibfnamefont {J.~C.~S.}\ \bibnamefont {Davis}},\ }\href
  {https://doi.org/10.1038/nature17411} {\bibfield  {journal} {\bibinfo
  {journal} {Nature}\ }\textbf {\bibinfo {volume} {532}},\ \bibinfo {pages}
  {343?347} (\bibinfo {year} {2016})}\BibitemShut {NoStop}%
\bibitem [{\citenamefont {Manning-Coe}\ and\ \citenamefont
  {Bradlyn}(2023)}]{bradlyn2023}%
  \BibitemOpen
  \bibfield  {author} {\bibinfo {author} {\bibfnamefont {D.}~\bibnamefont
  {Manning-Coe}}\ and\ \bibinfo {author} {\bibfnamefont {B.}~\bibnamefont
  {Bradlyn}},\ }\href {https://doi.org/10.1103/PhysRevB.108.165136} {\bibfield
  {journal} {\bibinfo  {journal} {Phys. Rev. B}\ }\textbf {\bibinfo {volume}
  {108}},\ \bibinfo {pages} {165136} (\bibinfo {year} {2023})}\BibitemShut
  {NoStop}%
\bibitem [{\citenamefont {Hackner}\ \emph {et~al.}(2025)\citenamefont
  {Hackner}, \citenamefont {Mai},\ and\ \citenamefont
  {Phillips}}]{hackner2025solving}%
  \BibitemOpen
  \bibfield  {author} {\bibinfo {author} {\bibfnamefont {N.~A.}\ \bibnamefont
  {Hackner}}, \bibinfo {author} {\bibfnamefont {P.}~\bibnamefont {Mai}},\ and\
  \bibinfo {author} {\bibfnamefont {P.~W.}\ \bibnamefont {Phillips}},\
  }\href@noop {} {\bibfield  {journal} {\bibinfo  {journal} {Physical Review
  A}\ }\textbf {\bibinfo {volume} {111}},\ \bibinfo {pages} {063316} (\bibinfo
  {year} {2025})}\BibitemShut {NoStop}%
\bibitem [{\citenamefont {Tenkila}\ \emph {et~al.}(2025)\citenamefont
  {Tenkila}, \citenamefont {Zhao},\ and\ \citenamefont
  {Phillips}}]{DynamicalPP2025}%
  \BibitemOpen
  \bibfield  {author} {\bibinfo {author} {\bibfnamefont {G.}~\bibnamefont
  {Tenkila}}, \bibinfo {author} {\bibfnamefont {J.}~\bibnamefont {Zhao}},\ and\
  \bibinfo {author} {\bibfnamefont {P.~W.}\ \bibnamefont {Phillips}},\ }\href
  {https://doi.org/10.1103/PhysRevB.111.045126} {\bibfield  {journal} {\bibinfo
   {journal} {Phys. Rev. B}\ }\textbf {\bibinfo {volume} {111}},\ \bibinfo
  {pages} {045126} (\bibinfo {year} {2025})}\BibitemShut {NoStop}%
\bibitem [{\citenamefont {Skolimowski}(2024)}]{realspace2024}%
  \BibitemOpen
  \bibfield  {author} {\bibinfo {author} {\bibfnamefont {J.}~\bibnamefont
  {Skolimowski}},\ }\href {https://doi.org/10.1103/PhysRevB.109.165129}
  {\bibfield  {journal} {\bibinfo  {journal} {Phys. Rev. B}\ }\textbf {\bibinfo
  {volume} {109}},\ \bibinfo {pages} {165129} (\bibinfo {year}
  {2024})}\BibitemShut {NoStop}%
\bibitem [{\citenamefont {Mai}\ \emph {et~al.}(2024)\citenamefont {Mai},
  \citenamefont {Zhao}, \citenamefont {Tenkila}, \citenamefont {Hackner},
  \citenamefont {Kush}, \citenamefont {Pan},\ and\ \citenamefont
  {Phillips}}]{mai2024newapproachstrongcorr}%
  \BibitemOpen
  \bibfield  {author} {\bibinfo {author} {\bibfnamefont {P.}~\bibnamefont
  {Mai}}, \bibinfo {author} {\bibfnamefont {J.}~\bibnamefont {Zhao}}, \bibinfo
  {author} {\bibfnamefont {G.}~\bibnamefont {Tenkila}}, \bibinfo {author}
  {\bibfnamefont {N.~A.}\ \bibnamefont {Hackner}}, \bibinfo {author}
  {\bibfnamefont {D.}~\bibnamefont {Kush}}, \bibinfo {author} {\bibfnamefont
  {D.}~\bibnamefont {Pan}},\ and\ \bibinfo {author} {\bibfnamefont {P.~W.}\
  \bibnamefont {Phillips}},\ }\href {https://arxiv.org/abs/2401.08746}
  {\bibinfo {title} {{New Approach to Strong Correlation: Twisting Hubbard into
  the Orbital Hatsugai-Kohmoto Model}}} (\bibinfo {year} {2024}),\ \Eprint
  {https://arxiv.org/abs/2401.08746} {arXiv:2401.08746 [cond-mat.str-el]}
  \BibitemShut {NoStop}%
\bibitem [{\citenamefont {Mai}\ \emph {et~al.}(2023{\natexlab{a}})\citenamefont
  {Mai}, \citenamefont {Zhao}, \citenamefont {Feldman},\ and\ \citenamefont
  {Phillips}}]{Mai_top_2023}%
  \BibitemOpen
  \bibfield  {author} {\bibinfo {author} {\bibfnamefont {P.}~\bibnamefont
  {Mai}}, \bibinfo {author} {\bibfnamefont {J.}~\bibnamefont {Zhao}}, \bibinfo
  {author} {\bibfnamefont {B.~E.}\ \bibnamefont {Feldman}},\ and\ \bibinfo
  {author} {\bibfnamefont {P.~W.}\ \bibnamefont {Phillips}},\ }\href
  {http://dx.doi.org/10.1038/s41467-023-41465-6} {\bibfield  {journal}
  {\bibinfo  {journal} {Nature Communications}\ }\textbf {\bibinfo {volume}
  {14}},\ \bibinfo {pages} {5999} (\bibinfo {year}
  {2023}{\natexlab{a}})}\BibitemShut {NoStop}%
\bibitem [{\citenamefont {Mai}\ \emph {et~al.}(2023{\natexlab{b}})\citenamefont
  {Mai}, \citenamefont {Feldman},\ and\ \citenamefont
  {Phillips}}]{Top_Mai_Haldane}%
  \BibitemOpen
  \bibfield  {author} {\bibinfo {author} {\bibfnamefont {P.}~\bibnamefont
  {Mai}}, \bibinfo {author} {\bibfnamefont {B.~E.}\ \bibnamefont {Feldman}},\
  and\ \bibinfo {author} {\bibfnamefont {P.~W.}\ \bibnamefont {Phillips}},\
  }\href {https://doi.org/10.1103/PhysRevResearch.5.013162} {\bibfield
  {journal} {\bibinfo  {journal} {Phys. Rev. Res.}\ }\textbf {\bibinfo {volume}
  {5}},\ \bibinfo {pages} {013162} (\bibinfo {year}
  {2023}{\natexlab{b}})}\BibitemShut {NoStop}%
\bibitem [{\citenamefont {Jab{\l}onowski}\ \emph {et~al.}(2023)\citenamefont
  {Jab{\l}onowski}, \citenamefont {Skolimowski}, \citenamefont {Brzezicki},
  \citenamefont {Byczuk},\ and\ \citenamefont
  {Wysoki{\'n}ski}}]{jablonowski2023topological}%
  \BibitemOpen
  \bibfield  {author} {\bibinfo {author} {\bibfnamefont {K.}~\bibnamefont
  {Jab{\l}onowski}}, \bibinfo {author} {\bibfnamefont {J.}~\bibnamefont
  {Skolimowski}}, \bibinfo {author} {\bibfnamefont {W.}~\bibnamefont
  {Brzezicki}}, \bibinfo {author} {\bibfnamefont {K.}~\bibnamefont {Byczuk}},\
  and\ \bibinfo {author} {\bibfnamefont {M.~M.}\ \bibnamefont
  {Wysoki{\'n}ski}},\ }\href@noop {} {\bibfield  {journal} {\bibinfo  {journal}
  {Physical Review B}\ }\textbf {\bibinfo {volume} {108}},\ \bibinfo {pages}
  {195145} (\bibinfo {year} {2023})}\BibitemShut {NoStop}%
\bibitem [{\citenamefont {Wysoki\ifmmode~\acute{n}\else \'{n}\fi{}ski}\ and\
  \citenamefont {Brzezicki}(2023)}]{spinorbit2023}%
  \BibitemOpen
  \bibfield  {author} {\bibinfo {author} {\bibfnamefont {M.~M.}\ \bibnamefont
  {Wysoki\ifmmode~\acute{n}\else \'{n}\fi{}ski}}\ and\ \bibinfo {author}
  {\bibfnamefont {W.}~\bibnamefont {Brzezicki}},\ }\href
  {https://doi.org/10.1103/PhysRevB.108.035121} {\bibfield  {journal} {\bibinfo
   {journal} {Phys. Rev. B}\ }\textbf {\bibinfo {volume} {108}},\ \bibinfo
  {pages} {035121} (\bibinfo {year} {2023})}\BibitemShut {NoStop}%
\bibitem [{\citenamefont {Mandal}\ and\ \citenamefont
  {Freire}(2021)}]{ips_hermann1}%
  \BibitemOpen
  \bibfield  {author} {\bibinfo {author} {\bibfnamefont {I.}~\bibnamefont
  {Mandal}}\ and\ \bibinfo {author} {\bibfnamefont {H.}~\bibnamefont
  {Freire}},\ }\href {https://doi.org/10.1103/PhysRevB.103.195116} {\bibfield
  {journal} {\bibinfo  {journal} {Phys. Rev. B}\ }\textbf {\bibinfo {volume}
  {103}},\ \bibinfo {pages} {195116} (\bibinfo {year} {2021})}\BibitemShut
  {NoStop}%
\bibitem [{\citenamefont {Freire}\ and\ \citenamefont
  {Mandal}(2021)}]{Freire_PLA_2021}%
  \BibitemOpen
  \bibfield  {author} {\bibinfo {author} {\bibfnamefont {H.}~\bibnamefont
  {Freire}}\ and\ \bibinfo {author} {\bibfnamefont {I.}~\bibnamefont
  {Mandal}},\ }\href {https://doi.org/10.1016/j.physleta.2021.127470}
  {\bibfield  {journal} {\bibinfo  {journal} {Physics Letters A}\ }\textbf
  {\bibinfo {volume} {407}},\ \bibinfo {pages} {127470} (\bibinfo {year}
  {2021})}\BibitemShut {NoStop}%
\bibitem [{\citenamefont {Mandal}\ and\ \citenamefont
  {Freire}(2024)}]{Hermann_review}%
  \BibitemOpen
  \bibfield  {author} {\bibinfo {author} {\bibfnamefont {I.}~\bibnamefont
  {Mandal}}\ and\ \bibinfo {author} {\bibfnamefont {H.}~\bibnamefont
  {Freire}},\ }\href {https://doi.org/10.1088/1361-648X/ad665e} {\bibfield
  {journal} {\bibinfo  {journal} {Journal of Physics: Condensed Matter}\
  }\textbf {\bibinfo {volume} {36}},\ \bibinfo {pages} {443002} (\bibinfo
  {year} {2024})}\BibitemShut {NoStop}%
\bibitem [{\citenamefont {Zhao}\ \emph {et~al.}(2023)\citenamefont {Zhao},
  \citenamefont {La~Nave},\ and\ \citenamefont {Phillips}}]{wicks}%
  \BibitemOpen
  \bibfield  {author} {\bibinfo {author} {\bibfnamefont {J.}~\bibnamefont
  {Zhao}}, \bibinfo {author} {\bibfnamefont {G.}~\bibnamefont {La~Nave}},\ and\
  \bibinfo {author} {\bibfnamefont {P.~W.}\ \bibnamefont {Phillips}},\ }\href
  {https://doi.org/10.1103/PhysRevB.108.165135} {\bibfield  {journal} {\bibinfo
   {journal} {Phys. Rev. B}\ }\textbf {\bibinfo {volume} {108}},\ \bibinfo
  {pages} {165135} (\bibinfo {year} {2023})}\BibitemShut {NoStop}%
\bibitem [{\citenamefont {Ma}\ \emph {et~al.}(2025)\citenamefont {Ma},
  \citenamefont {Zhao}, \citenamefont {Huang}, \citenamefont {Kush},
  \citenamefont {Bradlyn},\ and\ \citenamefont
  {Phillips}}]{ma2024chargesusceptibility}%
  \BibitemOpen
  \bibfield  {author} {\bibinfo {author} {\bibfnamefont {Y.}~\bibnamefont
  {Ma}}, \bibinfo {author} {\bibfnamefont {J.}~\bibnamefont {Zhao}}, \bibinfo
  {author} {\bibfnamefont {E.~W.}\ \bibnamefont {Huang}}, \bibinfo {author}
  {\bibfnamefont {D.}~\bibnamefont {Kush}}, \bibinfo {author} {\bibfnamefont
  {B.}~\bibnamefont {Bradlyn}},\ and\ \bibinfo {author} {\bibfnamefont {P.~W.}\
  \bibnamefont {Phillips}},\ }\href {https://doi.org/10.1103/r4q9-hm45}
  {\bibfield  {journal} {\bibinfo  {journal} {Phys. Rev. B}\ }\textbf {\bibinfo
  {volume} {112}},\ \bibinfo {pages} {045109} (\bibinfo {year}
  {2025})}\BibitemShut {NoStop}%
\end{thebibliography}

%

\end{document}